\documentclass[lettersize,journal]{IEEEtran}
\usepackage{amsmath,amsfonts}

\usepackage{array}
\usepackage{threeparttable}
\usepackage{subcaption}

\usepackage{textcomp}
\usepackage{booktabs}
\usepackage{stfloats}
\usepackage{url}
\usepackage{verbatim}
\usepackage{graphicx}
\usepackage{cite}
\usepackage{hyperref}
\usepackage[linesnumbered,ruled]{algorithm2e}
\begin{document}

\title{Patch-of-Interest ViT Inference Acceleration System for Edge-Assisted Video Analytics}

\author{Haosong~Peng,
         Wei~Feng,
         Hao~Li,
         Yufeng~Zhan,
         Ren~Jin,
         and~Yuanqing~Xia,~\IEEEmembership{Fellow,~IEEE}
         % and~Song~Guo,~\IEEEmembership{Fellow,~IEEE}% <-this % stops a space
 \IEEEcompsocitemizethanks{\IEEEcompsocthanksitem H.~Peng, Y.~Zhan, W.~Feng, and Y.~Xia are with the School of Automation, Beijing Institute of Technology, Beijing, China, 100086.
 E-mail: livion\_i@icloud.com~(Peng),~yu-feng.zhan@bit.edu.cn~(Zhan), 3220231069@bit.edu.cn~(Feng), xia\_yuanqing@bit.edu.cn~(Xia)
 \IEEEcompsocthanksitem R.~Jin is with the School of Aerospace Engineering, Beijing Institute of Technology, Beijing, China, 100086.
 E-mail: renjin@bit.edu.cn.
 \IEEEcompsocthanksitem H.~Li is with the School of Automation, Northwestern Polytechnology University, Xi'an, China, 710129.
 E-mail: lifugan\_10027@outlook.com
 % % \IEEEcompsocthanksitem P. Li is with the School of Computer Science and Engineering, the University of Aizu, Japan.\protect\\
 % % E-mail: pengli@u-aizu.ac.jp
 % \IEEEcompsocthanksitem X.~Zhang is with iF-Labs, Beijing Teleinfo Technology Co., Ltd., Beijing, China.
 % E-mail: zhangxiaopu@teleinfo.cn
 % \IEEEcompsocthanksitem S. Guo is with the Department of Computer Science and Engineering, The Hong Kong University of Science and Technology, Hong Kong SAR, China.
 % E-mail: songguo@cse.ust.hk
 }% <-this % stops an unwanted space
 % \thanks{This work has been submitted to the IEEE for possible publication. Copyright may be transferred without notice.}
 }


\markboth{Journal of \LaTeX\ Class Files,~Vol.~14, No.~8, August~2021}%
{Shell \MakeLowercase{\textit{et al.}}: A Sample Article Using IEEEtran.cls for IEEE Journals}


\maketitle

\begin{abstract}
The advent of edge computing has made real-time intelligent video analytics feasible. 
Previous works, based on traditional model architecture (e.g., CNN, RNN, etc.), employ various strategies to filter out non-region-of-interest content to minimize bandwidth and computation consumption but show inferior performance in adverse environments.
Recently, visual foundation models based on transformers have shown great performance in adverse environments due to their amazing generalization capability. 
However, they require a large amount of computation power, which limits their applications in real-time intelligent video analytics.
In this paper, we find visual foundation models like Vision Transformer (ViT) also have a dedicated acceleration mechanism for video analytics.
To this end, we introduce Arena, an end-to-end edge-assisted video inference acceleration system based on ViT.
We leverage the capability of ViT that can be accelerated through token pruning by only offloading and feeding Patches-of-Interest to the downstream models.
Additionally, we design an adaptive keyframe inference switching algorithm tailored to different videos, capable of adapting to the current video content to jointly optimize accuracy and bandwidth. 
Through extensive experiments, our findings reveal that Arena can boost inference speeds by up to 1.58\(\times\) and 1.82\(\times\) on average while consuming only 47\% and 31\% of the bandwidth, respectively, all with high inference accuracy.
\end{abstract}

\begin{IEEEkeywords}
Video Analytics System, Edge Computing, Vision Transformer.
\end{IEEEkeywords}

\section{Introduction}
Today, IoT cameras are generating an unprecedented volume of video data daily.
According to predictions by Hailo~\cite{Hailo}, the global market is anticipated to see approximately 180 million shipments of embedded video surveillance devices by 2025. 
These cameras power a plethora of video analytics applications such as traffic management~\cite{li2023cross},
% ,li2023cross
autonomous driving~\cite{zhang2021emp},
% ,zhang2021emp
security surveillance~\cite{yi2020eagleeye}, etc. 
% ,yi2020eagleeye

The massive data generated by IoT cameras far exceeds their limited computing capabilities. 
A straightforward way is to offload the captured data to edge servers~\cite{du2020server,zeng2020distream,lv2023feedback,liu2022adamask,yang2022edgeduet,zhang2021elf,tchaye2022smartfilter,gao2022ivp,zhang2024novas,sabet2021similarity}.
These servers execute intensive downstream inference tasks, usually based on deep neural networks, and deliver useful visual feedback to users.
Previous works have explored the spatial and temporal redundancy of video content to design systems that reduce the bandwidth usage between cameras and edge servers. 
Efforts are made either to extract and transmit regions of interest (RoIs) from video frames~\cite{liu2022adamask,zhang2021elf} or to design a module that filters out similar frames~\cite{tchaye2022smartfilter,chen2015glimpse}.  
Though these methods have optimized the transmission and computation, their downstream models based on Convolutional Neural Network (CNN) show inferior performance in the live video data, because of the poor generalization ability of CNN.
\begin{figure}[!t]
\begin{center}
\includegraphics[width=1\linewidth]{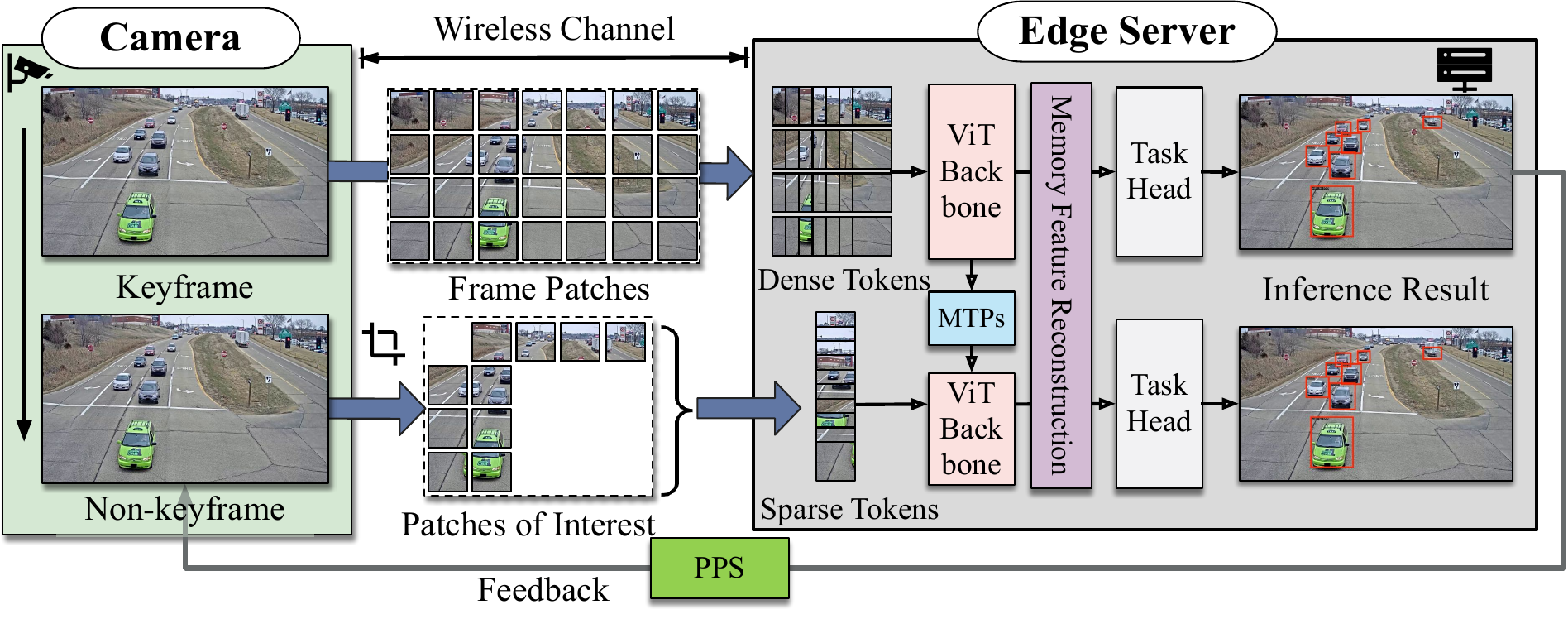}
\caption{Arena: our patch-of-interest ViT inference acceleration system for edge-assisted video analytics. Due to the limited computing power of the camera, the extracted patches-of-interest are offloaded to an edge server for processing with its more powerful GPUs. MTPs stands for Memory Token Pools.}
\label{fig:intro}
\end{center}
\end{figure}

In recent years, visual foundation models based on transformer architecture~\cite{dosovitskiy2020image,Video_swin_transformer,he2022masked}, exemplified by Vision Transformer (ViT)~\cite{dosovitskiy2020image}, have surpassed CNN-based methods
% ~\cite{girshick2015fast,redmon2018yolov3,Cai_2019,tian2019fcos} 
in most downstream tasks, including image classification,
%~\cite{deng2009imagenet}%
object detection
% ~\cite{lin2014microsoft} 
and object tracking.
% ~\cite{milan2016mot16}
%
The strengths of the ViT lie in its exceptional feature extraction capabilities and generalization. 
Through its multi-head self-attention mechanism (MSA), ViT can more effectively capture global information within the input, achieving superior performance in various aspects.
Furthermore, the ``pre-training + fine-tuning" paradigm has also standardized the implementation process of video analytics, enabling models to be rapidly deployed in entirely new specific scenes using only a few samples.
However, the computational demands of its MSA increase quadratically with the number of patches.
This intense computational requirement, especially in high-resolution video (e.g., 1080p)  analytics scenarios, impedes the practical application deployment of ViT-based models.
Although there exits works on efficient ViT that utilize token pruning~\cite{rao2021dynamicvit,liang2022not}, merging~\cite{bolya2022token,zeng2022not}, and fusion~\cite{kim2024token} techniques to reduce computation, these approaches still require a full image as input and fail to adapt to real-time intelligent video analytics.

% complex dense prediction downstream tasks.

To fill this gap, we propose Arena, a Patch-of-Interest (PoI) ViT inference acceleration system for edge-assisted video analytics. 
We find that ViT also has a dedicated acceleration mechanism for video analytics.
As shown in Figure~\ref{fig:intro}, first, the camera transmits a full video frame (referred to as a keyframe) to the edge server, followed by conducting inference on it, where we establish and maintain memory token pools.
Subsequently, we utilize a Probability-based Patch Sampling (PPS) mechanism to determine the PoIs for the next few frames (referred to as a Non-keyframe). 
Therefore, only PoIs, usually occupying a small portion of the frame, are required to transmit to the edge server.
Then, a model based on the ViT backbone performs inference merely using these PoIs.
Thus, inference on non-keyframes significantly reduces the workload compared to keyframe inference.
Finally, a Memory Feature Reconstruction (MFR) module is employed to restore the complete feature maps, making it available for a dense prediction task head.

Arena periodically alternates between the keyframe inference and non-keyframe inference phases.
Due to the temporal and spatial characteristics of different cameras, the amount and timing of the appearance of task-relevant objects vary as described in~\cite{peng2024tangram}. 
For example, on a street at night, the number of interesting objects, such as people and cars, is much fewer compared to daytime. 
Similarly, the number of objects in remote rural videos differs significantly from that in urban areas. 
Therefore, a fixed strategy for switching between keyframes and non-keyframes is not suitable for all video analytics scenarios. 
Consequently, we propose an Adaptive Keyframe Interval Switching (AKIS) algorithm to control when to switch between keyframe inference and non-keyframe inference.

Arena enhances overall system throughput and computational efficiency twofold. 
On the one hand, transmitting only partial patches significantly reduces bandwidth consumption and transmission latency. 
On the other hand, feeding PoIs to the ViT backbone and reconstructing features from memory tokens reduce computational overhead and inference time while maintaining almost the same accuracy.
Additionally, the AKIS algorithm can personalize the configuration of the keyframe interval for each video, further achieving a trade-off between accuracy and bandwidth.
We develop and deploy a prototype on a testbed that uses an NVIDIA Jetson as a camera device and an Ubuntu desktop as an edge server running real video analytics workloads.
We summarize the contributions of this work as follows:
\begin{itemize}
    \item We propose Arena, an edge-assisted video analytics system that utilizes PoI ViT inference, accelerating inference while reducing network bandwidth. 
    \item  We have designed an adaptive keyframe interval switching algorithm that adapts to the current video content, jointly optimizing accuracy and bandwidth in real-time.
    \item Experimental results on the testbed show that Arena can accelerate the inference up to $1.58\times$ and $1.82\times$ on average by only using 47\% and 31\% bandwidth while keeping the inference accuracy sacrifice within an acceptance range.
\end{itemize}

The remainder of this paper is organized as follows. 
We present the preliminaries and motivation in Section~\ref{sec_II}. 
The design of Arena is introduced in Section~\ref{sec_III}, followed by the system implementation in Section~\ref{sec_IV}. 
We conduct extensive experiments in Section~\ref{sec_V}, and the related work is reviewed in Section~\ref{sec_VI}. 
Finally, we conclude this paper in Section~\ref{sec_VII}.

\section{Preliminaries And Motivation}\label{sec_II}
We first introduce the preliminaries of Vision Transformer (Sec.~\ref{sec:pre-vit}).
Next, we reveal that video data contains a significant amount of redundancy. 
Substantial bandwidth can be saved by only transmitting the PoIs (Sec.~\ref{sec:redundancy}). 
However, traditional CNN-based models cannot efficiently process these unstructured regions (Sec.~\ref{sec:problem}).
In contrast, ViT is well-suited for unstructured data because it can process variable-length sequences as inputs. 
This capability allows for filtering out data at the beginning of the pipeline and accelerates the computation (Sec.~\ref{sec:reduction}), which gives us insights on how to accelerate ViT-based video analytics systems.
Finally, the varying trade-offs induced by keyframe intervals across different scenarios (Sec.~\ref{sec:tradeoff}) inspire us to develop methods for keyframe interval adjustment.

\subsection{Preliminaries of Vision Transformer}
\label{sec:pre-vit}
% Despite the limited receptive field in CNN-based detectors to capture local spatial contexts from the image, ViT excels at integrating global spatial contexts across entire images, thanks to its attention mechanisms.
%
Given an input image $\mathbf{x} \in \mathbb{R}^{H \times W \times C}$, ViT splits it into a sequence of flattened 2D patches $\mathbf{x}_p=\{x_{p,1}, \cdots, x_{p,N}| x_{p,i} \in \mathbb{R}^{P^2 C}\} $.
Here, $H\times W$ represents the resolution of the frame, $C$ is the number of channels, \(P\) denotes the size of each patch, and $N = (H\times W) / P^2$ is the number of patches after splitting.
After that, ViT constructs initial tokens as 
\begin{equation}
\tilde{z}_{0} =\left[x_{class}, x_{p, 1} \mathbf{E} ; x_{p, 2} \mathbf{E} ; \ldots ; x_{p, N} \mathbf{E}\right], 
\label{eq:vit1}
\end{equation}
where \(x_{class}\) is a learnable embedding that prepends into flattened sequences and \(\mathbf{E} \in \mathbb{R}^{P^2  C \times D}\) represents the parameters of the linear projection, and $D$ denotes the embedding dimension of each token.
Moreover, ViT adds position embeddings \(\mathbf{E}_{pos} \) on \(\tilde{z}_{0}\) to retain its positional awareness as
\begin{equation}
     z_0 = \tilde{z}_{0} + \mathbf{E}_{pos}, \qquad \mathbf{E}_{pos}\in \mathbb{R}^{(N+1)\times D}.
\label{eq:vit2}
\end{equation}
In the end, the transformer encoder consisting of MSA, MLP blocks, and LayerNorm (LN) is adopted to update \(z_0\) with \(L\) iteration and output the final image representation \(\mathbf{y}\). The formulations are
\begin{equation}
\begin{aligned}
z_{\ell}^{\prime} & =\operatorname{MSA}\left(\operatorname{LN}(z_{\ell-1})\right) + z_{\ell-1}^{\prime}, & & \ell=1 \ldots L, \\
z_{\ell} & =\operatorname{MLP}\left(\operatorname{LN}(z^{\prime}{ }_{\ell})\right)+z_{\ell}^{\prime}, & & \ell=1 \ldots L, 
\end{aligned}
\label{eq:vit3}
\end{equation}

\begin{equation}
\mathbf{y}=\mathrm{LN}(z_L[0]).
\end{equation}

\subsection{Redundancy in Video Inference Data}
\label{sec:redundancy}
In edge-assisted video analytics, high-resolution videos often include much redundant data. 
%
% Taking the object detection as an example, a small region containing objects within each video frame is identified as an RoI.
%
Particularly, the video frame is divided into multiple patches of certain pixels (e.g., $16\times 16$), among which those containing the objects are referred to as PoIs.
Conversely, the rest of the frame is dominated by the background (e.g., buildings and sky) and other irrelevant objects~\cite{jiang2021flexible}.
Table~\ref{tab:table1} shows the redundancy of the first five cameras from two popular real-world video datasets. 
It is evident that in most videos, the proportion of PoI does not exceed 25\%.
Additionally, in some specific scenes, PoIs constitute no more than 5\% of the whole frame.
By only transmitting the PoIs from the camera to the edge server, most videos can achieve a bandwidth usage reduction exceeding 80\%, with potential savings reaching up to 92.01\%.
This discovery motivates us to seek a more fine-grained method for extracting PoIs that allows transmitting as few patches as possible while still meeting the accuracy requirements.
%Feeding only the PoIs from video frames into the ViT-small backbone for inference yields two substantial advantages.
%
%On one hand, most videos can achieve a latency reduction of over 80\%, with a maximum savings of up to 90.11\%
%
%On the other hand, it can also save bandwidth consumption up to 92.01\%.
%
%In summary, video analysis commonly contains large redundancies, not only increasing bandwidth consumption but also leading to inefficient video inference.

\begin{table}[!]
\caption{Redundancy in video inference data on MOT17Det~\cite{milan2016mot16} and AIC22~\cite{Naphade22AIC22} datasets.}
\resizebox{0.48\textwidth}{!}{
\begin{tabular}{@{}c|c|c|c|c@{}}
\toprule
\textbf{Scene Name (\# Frame)} & \textbf{\# Object} & \textbf{PoI Prop.$^\triangle$} & \textbf{Latency$^\diamondsuit$} & \textbf{Bandwidth}$^\star$ \\ \midrule
MOT17-02 (600) & 9393 & 17.10\% & 90.11\%  & 67.29\% \\
MOT17-04  (1050) & 39821 & 22.59\% & 86.55\% & 62.60\%  \\
MOT17-05 (837) & 4767 & 69.21\% & 48.01\% & 21.70\% \\
MOT17-09 (525) & 3745 & 24.46\% & 85.67\% & 57.72\% \\
MOT17-10 (654) & 10560 & 13.76\% & 89.14\%& 56.64\% \\ \midrule
AIC22-c001 (1955) & 3310 & 3.79\% & 86.76\% & 92.01\% \\
AIC22-c002 (2110) & 6322 & 4.77\% & 85.24\% & 90.86\% \\
AIC22-c003 (1996) & 6677 & 8.69\% & 85.40\% & 87.14\% \\
AIC22-c004 (2110) & 6015 & 4.88\% & 84.92\% & 88.05\% \\
AIC22-c005 (2110) & 12199 & 5.40\% & 74.60\% & 88.75\% \\ \bottomrule
\end{tabular}
}
\begin{tablenotes}
\footnotesize
\item \# represents “The number of”;
\item $\triangle$ The average proportion of the total PoIs in each frame;  %此处加入
\item $\diamondsuit$ Average proportion of inference latency saved by only computing PoIs on ViT backbone; 
\item $\star$ Saved bandwidth proportion if only transmits the PoIs.
%此处加入注释**信息
\end{tablenotes}
\label{tab:table1}
\end{table}

\subsection{The Limitation of CNN-based Backbone on Unstructured Input}\label{sec:problem}

\begin{figure}[!t]
\begin{center}
\includegraphics[width=1\linewidth]{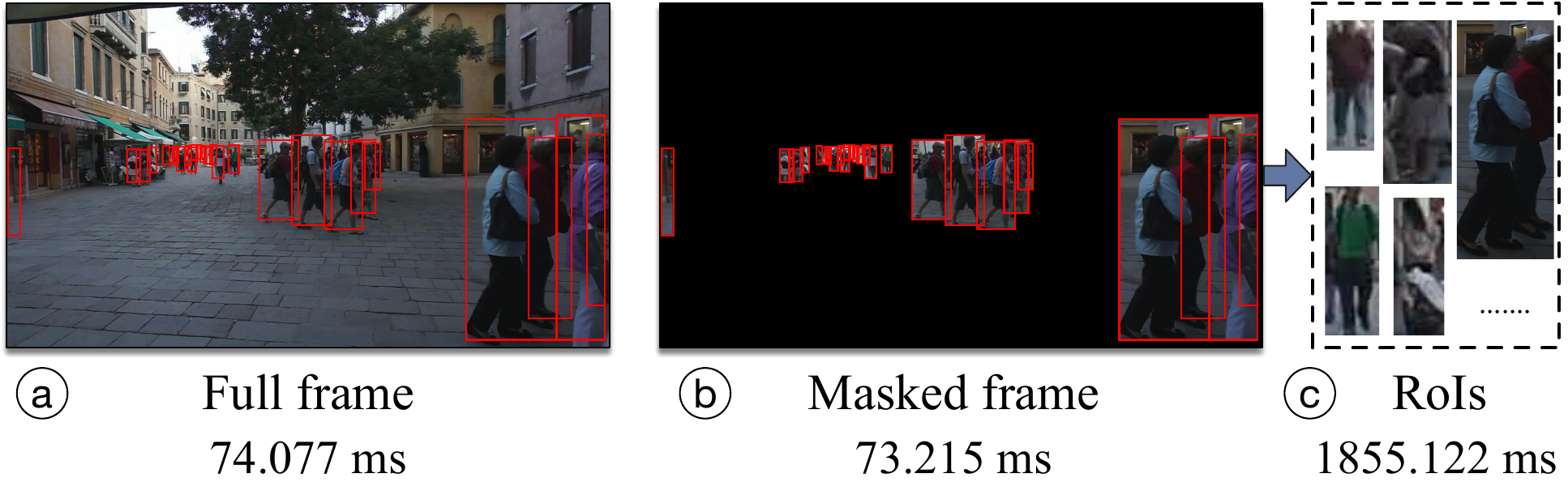}
\caption{The inference latency for three strategies: Full Frame, Masked Frame, and RoIs separately. Downstream models based on CNNs fail to benefit from the filtered RoIs. }
\label{fig:limitation}
\end{center}
\end{figure}

Models based on a CNN backbone struggle to handle unstructured inputs. 
Previous approaches use lightweight methods to filter RoI from the video frame~\cite{du2020server,liu2022adamask,li2020reducto,cheng2023edge}, which is indeed effective in reducing bandwidth consumption. 
However, they either infer masked, full-size images or process each RoI individually.
The former does not reduce the computation on the backbone because of the same size, while the latter requires multiple inferences and cannot batch together due to varying RoI sizes. 
For example, we follow the strategy from~\cite{du2020server} to extract these regions whose objectness scores are greater than 50\%, as shown in Figure~\ref{fig:limitation}a. 
Figures~\ref{fig:limitation}b and \ref{fig:limitation}c illustrate the masked image and RoI respectively. 
We employ a full detector to infer these two types of unstructured data, with latencies of 73.215ms and 1855.122ms, respectively, which do not achieve acceleration compared to the original image's inference time of 74.077ms. 
Therefore, downstream models based on CNNs fail to achieve acceleration benefits from the filtered RoIs.

\subsection{Accelerating ViT via Patch Prunning}\label{sec:reduction}

\begin{figure}[!t]
\begin{center}
\includegraphics[width=1\linewidth]{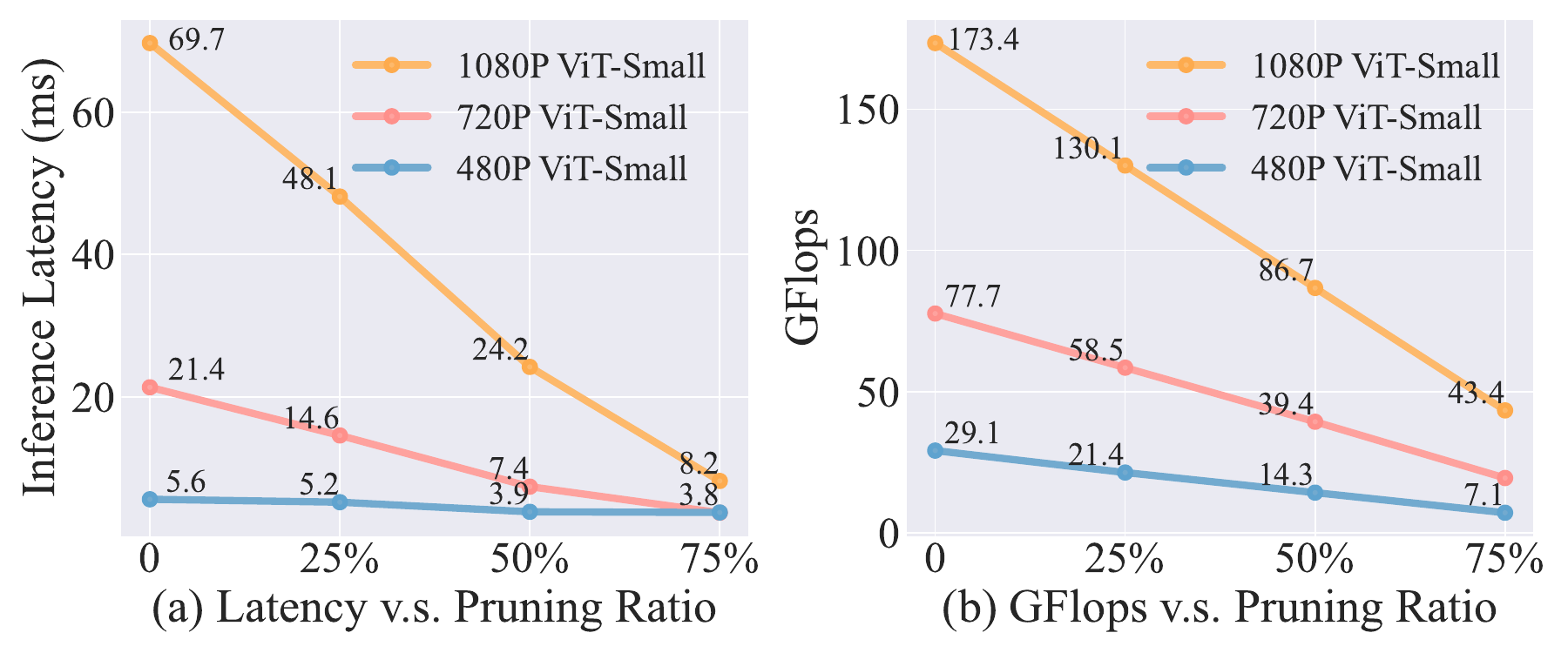}
\caption{Pruning patches can accelerate ViT backbone inference. We evaluate the impact of pruning 25\%, 50\%, and 75\% tokens on~(a) inference latency and~(b) GFlops using ViT-base across videos of 1080p, 720p, and 480p resolutions. Similar trends were found in ViT-Base.}
\label{fig:motivation}
\end{center}
\end{figure}

One of the desirable advantages of ViT is its ability to allow a flexible length of patch input.
%
%Specifically, let $D$ be the embedding dimension of each token.
%
Specifically, the total computational complexity of one MSA module and one MLP is $\mathcal{O}\left(12 N D^2+2 N^2 D\right)$. 
Obviously, reducing the number of patches can linearly or even quadratically reduce the operations.
Figure~\ref{fig:motivation} illustrates the (a) average inference latency and (b) average GFlops per frame when different proportions of patches are pruned across videos of various resolutions.
It is noteworthy that using ViT-small on 1080p videos and pruning 75\% patches resulted in a reduction of latency by 62ms, only 11.76\% of the original latency.
Additionally, the GFlops are linearly related to the number of input patches, showing a reduction of approximately 74.97\% if pruning 75\% patches.
Furthermore, we confirm this on video datasets in Table~\ref{tab:table1}. 
The \textbf{Latency$^\diamondsuit$} column reports the average inference latency saved by only feeding the PoIs to the ViT backbone for each frame.
Most videos can achieve over an 80\% acceleration, with the highest reaching up to 90.11\%.
This observation shows the potential opportunities for accelerating ViT inference by pruning the number of input patches.

However, naively pruning patches is not feasible for dense prediction tasks, such as object detection, since they typically require full-sized feature maps.
How to adapt to downstream models while enjoying the inference acceleration brought by patch pruning presents a significant challenge to the design of Arena.
\begin{figure}[!t]
\begin{center}
\includegraphics[width=1\linewidth]{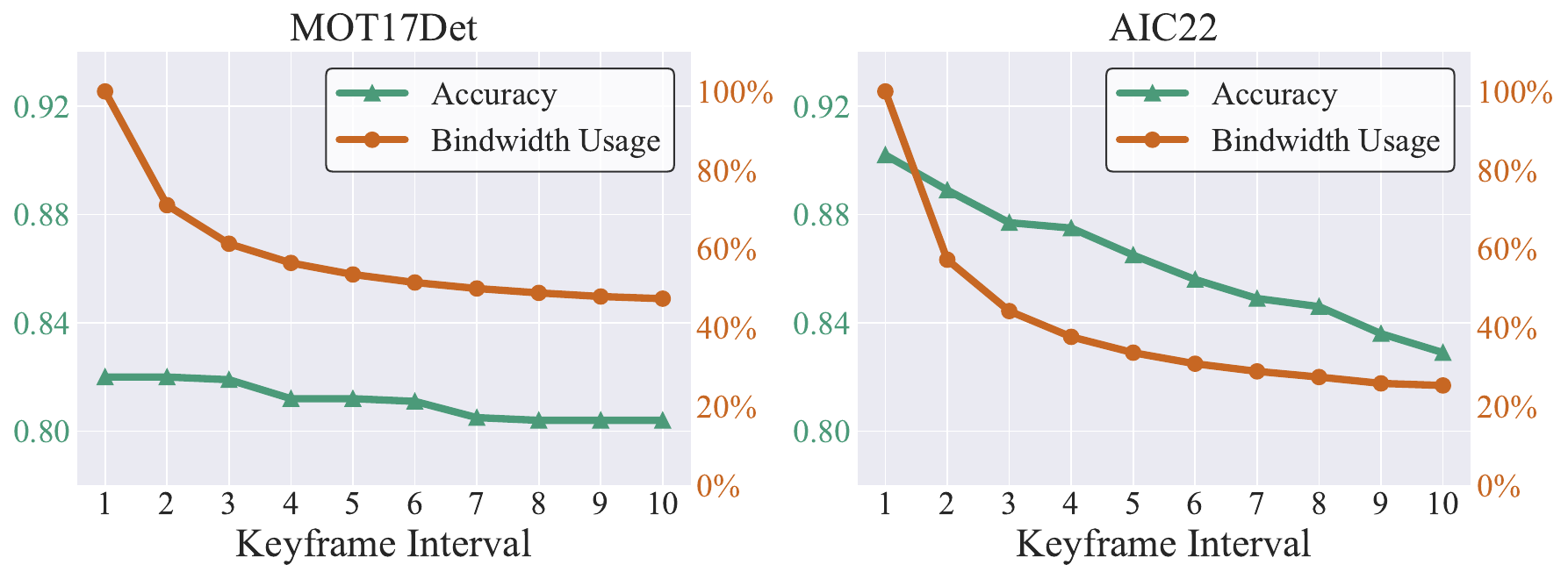}
\caption{Keyframe interval introduces accuracy and bandwidth trade-off in different scenes.}
\label{fig:keyframe_interval}
\end{center}
\vspace{-5mm}
\end{figure}

\subsection{
% The inference difficulty of various video scenes is influenced by both the frame rate at which the camera samples and the specific characteristics of the scene
Keyframe Interval Introduces Accuracy and Bandwidth Trade-off
}\label{sec:tradeoff}

\begin{comment}
%这里写视频画面变化大
In practical video analysis systems, while the key-frame offloading inference paradigm is an effective method for accelerating edge inference, the non-linear changes between actual frames result in varying levels of inference difficulty across different time intervals.
%
So the determination of the optimal keyframe interval is of paramount importance.
%
A naive keyframe scheduling policy picks a keyframe at a pre-fixed rate (e.g., every $K$ frames)\cite{zhu2017deep}.

We conduct experiments using different intervals with naive keyframe scheduling policy across various scenarios to analyze the trade-offs they introduce.
%
As Fig.? shows,...
%
Different keyframe intervals present distinct trade-offs across various scenarios.
%
Therefore,  scenarios in which the keyframe interval increases without a notable alteration in accuracy can be classified as ``easy" scenarios, which indicates that the features remain relatively consistent between adjacent frames.
%
In contrast, scenarios in which a gradual increase in keyframe interval results in a notable decline in accuracy can be classified as ``hard" scenarios.
\end{comment}
Cameras are commonly deployed in dynamic scenes such as traffic intersections, building entrances, and pedestrian streets, where the quantity of PoIs changes frequently.
A naive keyframe scheduling policy picks a keyframe at a pre-fixed rate (e.g., every $K$ frame~\cite{zhu2017deep}) is not feasible for different scenes.

We utilize two distinct types of video datasets and apply fixed keyframe interval strategies with varying intervals. 
MOT17Det is a dataset containing pedestrians on moving or stationary streets, where the number of objects in each frame is small, and their movement speed is slow. 
In contrast, AIC22 is a traffic dataset comprising moving vehicles, with a high number of objects moving at fast speeds.
As illustrated in Fig.~\ref{fig:keyframe_interval}, increasing the keyframe interval in MOT17Det has a negligible impact on accuracy, while significantly reducing bandwidth consumption. 
However, in AIC22, although increasing the keyframe interval similarly reduces bandwidth consumption, it also leads to a linear decrease in accuracy.
Different keyframe intervals thus present distinct trade-offs across various datasets. 
This phenomenon is observed not only across different datasets with varying objects but can also occur across different videos in the same dataset, as the moving speed of objects in videos is constantly changing.
% %
% Different keyframe intervals present distinct trade-offs across various datasets.
% %
% This phenomenon is not only observed across different datasets with different objects, but it can also occur across different videos in the same dataset because the moving speed of objects in videos is constantly changing.

The above experiments demonstrate that a fixed keyframe interval is not suitable for videos with dynamic changes.
Designing a strategy that adaptively adjusts the keyframe interval to balance accuracy and bandwidth in a real-time video analytics system is a crucial challenge for Arena.

\section{Arena Design}\label{sec_III}
\begin{figure*}[!t]
\begin{center}
\includegraphics[width=1\linewidth]{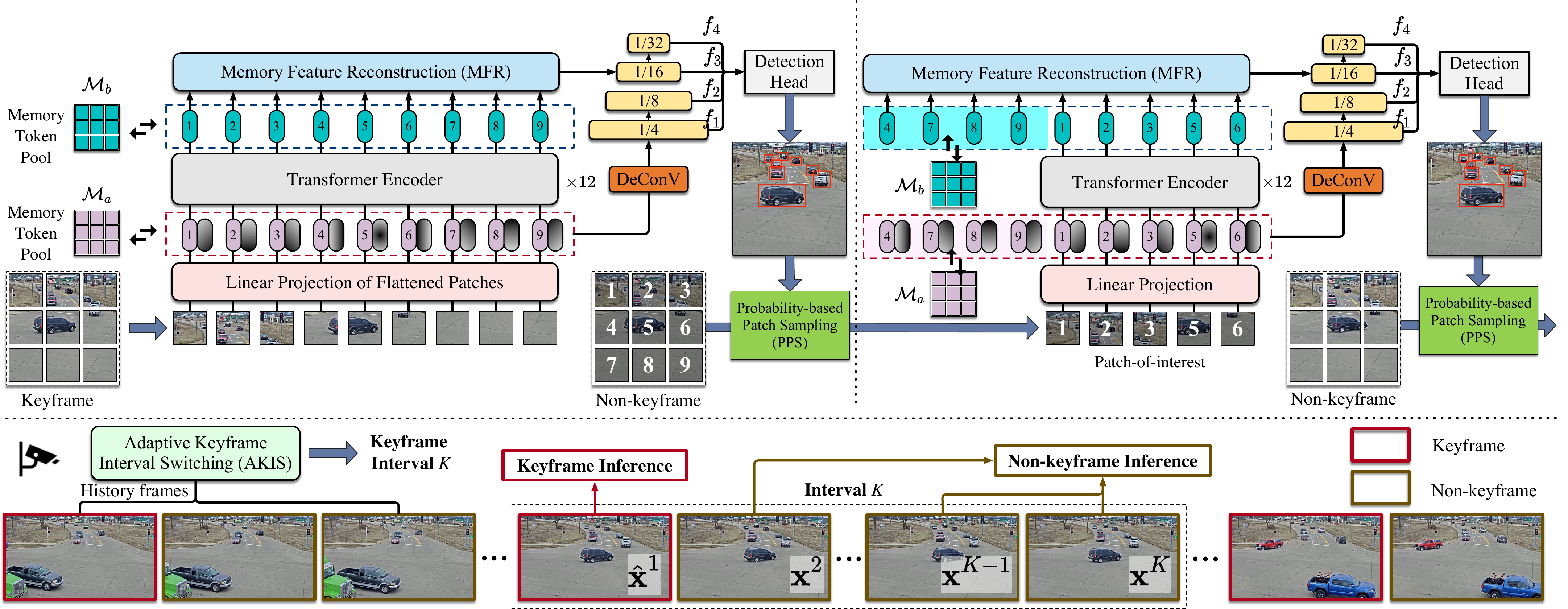}
\caption{The overview of Arena. Given \(K\) continuous frames  \(\{\hat{\mathbf{x}}^1, \mathbf{x}^2, \ldots, \mathbf{x}^K\}\) in an interval, Arena periodically operates in two distinct phases: keyframe inference (Left) for the first frame \(\hat{\mathbf{x}}^1\) and non-keyframe inference (Right) for the rest of the frames. AKIS (Down), deployed on the camera, utilizes information from historical frames to determine subsequent keyframe intervals.  Notably, we split the frame into nine patches only for demonstration. }
\label{fig:overview}
\end{center}
\vspace{-3mm}
\end{figure*}

\subsection{Overview}
Arena is an edge-assisted, real-time inference acceleration system tailored for ViT-based models. 
It aims to reduce communication and computation overheads by pruning the number of patches offloaded to the server while maintaining high accuracy.
%accelerate video analytics and decrease the computational overhead by reducing the number of patches offloaded to the server while maintaining high accuracy.
%
Arena comprises a camera and an edge server. 
The camera, equipped with limited computational power, continuously captures video frames, and the edge server deploys a ViT-based detector to perform inference.
In every $K$ frames, the first frame is designated as the keyframe, and the subsequent $K-1$ frames are considered as non-keyframes. 
Arena operates in two distinct phases: keyframe inference (Sec.~\ref{subsec_I}) and non-keyframe inference (Sec.~\ref{subsec_II}), then alternating periodically between them as illustrated in Figure~\ref{fig:overview}.  
In keyframe inference, Arena first performs full frame inference by detector and caches intermediate tokens into two memory token pools.
In the non-keyframe inference phase, it queries memory tokens from the pools and reconstructs complete image-wise features to accommodate unstructured samples with variable-length patch sequences.
Moreover, we introduce PPS to identify the PoIs in non-keyframes and provide feedback to the camera (Sec.~\ref{subsec_III}).
Additionally, the AKIS algorithm is proposed to personalize the settings for keyframe and non-keyframe intervals, further optimizing the trade-off between accuracy and bandwidth (Sec.~\ref{subsec_IV}).
Finally, Sec.~\ref{subsec:train} describes our training approach in a video-training manner.
Notably, both phases employ the same network and shared weights.
We take a sequence of frames \(\{\hat{\mathbf{x}}^1, \mathbf{x}^2, \ldots, \mathbf{x}^K\}\) as an example, where \(\hat{\mathbf{x}}^1\) denotes the keyframe and \(\mathbf{x}^2, \ldots, \mathbf{x}^K\) as non-keyframes, and detail our above-mentioned mechanisms in the following paragraphs.

\subsection{Keyframe Inference}\label{subsec_I}
In the keyframe inference phase, the camera transmits a full frame to the edge server.
%
%Additionally, the first frame of each scene from the camera is also treated as a keyframe.
For keyframe \(\hat{\mathbf{x}}^1\), the ViT detector splits it into a sequence of non-overlapping patches with resolution $P \times P$, denoted as $\hat{\mathbf{x}}_p^1 = \{\hat{x}^1_{p,1}, \cdots, \hat{x}^1_{p,N}| \hat{x}^1_{p,i} \in \mathbb{R}^{P^2C}\}$. 
%
%Here, $H\times W$ represents the resolution of the original frame, $C$ is the number of channels, and $N = H\times W / P^2$ is the number of patches after splitting.
%
After that, following the procedure of ViT, Arena constructs the initial tokens \(\hat{\tilde{z}}_0^{1}\) as
\begin{equation}
\mathcal{M}_a\leftarrow \hat{\tilde{z}}_0^{1} =[\hat{x}^1_{p,1} \mathbf{E} ; \hat{x}^1_{p,2} \mathbf{E} ; \ldots ; \hat{x}^1_{p,N} \mathbf{E}],
\end{equation}
where the initial tokens \(\hat{\tilde{z}}_0^{1}\) without positional encoding will be cached in the memory token pool \(\mathcal{M}_a\). 
After that, position embeddings \(\mathbf{E}_{pos}\) are added to the initial tokens to retain positional information as
\begin{equation}
\hat{z}_0^{1} =\hat{\tilde{z}}_0^{1} + \mathbf{E}_{pos}, \qquad \mathbf{E}_{pos}\in \mathbb{R}^{N\times D}.
\end{equation}
Then, all tokens are processed through $L$ transformer blocks to get the final output~\(\hat{z}_{L}^{1}\), which will be cached in the second memory token pool \(\mathcal{M}_b\). 
The formation is
\begin{equation}\label{encoder}
\begin{aligned}
        \hat{z}_{\ell}^{1\prime} & =\operatorname{MSA}\left(\operatorname{LN}(\hat{z}^{1}_{\ell-1})\right)+\hat{z}^{1}_{\ell-1}, & & \ell=1 \ldots L, \\
\hat{z}_{\ell}^1 & =\operatorname{MLP}\left(\operatorname{LN}(\hat{z}_{\ell}^{1\prime})\right)+\hat{z}_{\ell}^{1\prime}, & & \ell=1 \ldots L, 
\end{aligned}
\end{equation}
\begin{equation}
\mathcal{M}_b \leftarrow \hat{z}_L^1.
\end{equation}

Notably, memory token pools $\mathcal{M}_a$ and $\mathcal{M}_b$ will be updated in every frame inference and re-initialized at the next keyframe inference phase. 
Since the tokens of the keyframe are complete, we directly feed $\hat{z}_L^1$ into the MFR module. 
MFR is a single-layer transformer decoder~\cite{vaswani2017attention} to build the final tokens and reshape to the strongest feature map $f_3$.

We build a multi-scale feature pyramid \(\{f_1,f_2,f_3,f_4\}\) directly from intermediate tokens.
$f_1$ and $f_2$ contain low-level information, while $f_3$ and $f_4$, after processing through the transformer encoder, contain high-level semantic information.
Therefore, we combine these two sets of feature maps to enhance the effectiveness of object detection.
Specifically, we use the token sequence before encoder (i.e., $\hat{\tilde{z}}_0^{1}$) and after MFR (i.e., $f_3$) as features, which are both at $1/16$ scale relative to the original frame. 
For $\hat{\tilde{z}}_0^{1}$, we use two sets of de-convolution operations to obtain feature maps $f_1$, $f_2$ at $1/4$ and $1/8$ scale. 
For~$f_3$, we use the convolution of stride 2 to generate a feature map $f_4$ at $1/32$ scale.
The formulations are 
\begin{equation}\label{deconv}
f_1=\text{DeConV}_1(\hat{\tilde{z}}_0^1),\qquad f_2=\text{DeConV}_2(\hat{\tilde{z}}_0^1),
\end{equation}
\begin{equation}\label{conv}
f_3 = \text{MFR}(\hat{z}^1_L),\qquad f_4=\text{ConV}(f_3).
\end{equation}
Finally, $\{f_1, f_2, f_3, f_4\}$ from the ViT backbone form a multi-scale feature pyramid, which is then fed into two-stage detectors, such as Faster R-CNN \cite{ren2015faster} or Cascade R-CNN \cite{Cai_2019}, to generate the bounding boxes and categories of the objects.

\subsection{Non-keyframe Inference with Memory Feature Reconstruction}\label{subsec_II}

During non-keyframe inference, the PPS first identifies $N^\prime$ PoIs using the last frame detection results~(e.g.,~results from $\hat{\mathbf x}^1)$.
Then, the camera only extracts the PoIs out of the frame and offloads them to the edge server, which is denoted as $\mathbf{x}_p^k = \{x^k_{p,1}, \ldots, x^k_{p,N'}| x^k_{p,i} \in \mathbb{R}^{P^2 C}, k \in (1,K], i \in [1,N']\}$. 
%
%The edge server stores memory token pool \(\mathcal{M}^{t-1}_a\) and \(\mathcal{M}^{t-1}_b\) from \(t-1\) time step.
%
Here $N^\prime \ll N$ due to the fact that objects only occupy a very small portion of the frame.
%
%$\mathbf{x}^{t+1}$ is input to the same linear projection to obtain a new $z_0^{t+1\prime}$ at time step $t+1$.
After linear projection for $\mathbf{x}_p^k$, we get a sparse token sequence $\tilde{z}_{0,sp}^{k}$.
Subsequently, we add position embeddings corresponding to the original frame position to each token as
% before feeding them into the transformer encoder as
\begin{equation}
z_{0,sp}^{k} =\tilde{z}_{0,sp}^{k} + \mathbf{E}^\prime_{pos,sp}, \qquad \mathbf{E}^\prime_{pos,sp}\in \mathbb{R}^{N^\prime\times D}.
\end{equation}
Next, $z_{0,sp}^{k}$ is fed into the transformer encoder to compute~$z_{L,sp}^{k}$ following Eq.~(\ref{encoder}). 

To restore the complete token sequence from the sparse one,
we need to reuse the tokens in memory token pools.
We extract tokens from non-PoI regions in $\mathcal{M}_a$ and combine them with the PoIs token $\tilde{z}_{0,sp}^{k}$ to restore a complete token sequence $\tilde{z}_{0}^{k}$.
Similarly, we extract tokens from non-PoI regions in $\mathcal{M}_b$ and combine them with $z_{L,sp}^{k}$ to restore $z_{L}^{k}$.
For example, as illustrated in Figure~\ref{fig:overview}, the edge server only receives and processes tokens at positions 1, 2, 3, 5, and 6. 
Then, we insert the tokens from the remaining positions of \(\mathcal{M}_a\) and \(\mathcal{M}_b\) (i.e., 4, 7, 8, and 9) into the processed tokens $\tilde{z}_{0,sp}^{k}$ and $z_{L,sp}^{k}$ to restore complete sequences, respectively.
Meanwhile, we introduce MFR to enhance the PoI tokens with background information from historical frames with minimal computational overhead.
Finally, we follow the Eq.~(\ref{deconv}) and Eq.~(\ref{conv}) to get the multi-scale feature pyramid \(\{f_1,f_2,f_3,f_4\}\). 

\begin{comment}
For feature maps $f_1$ and $f_2$, we fill tokens $\tilde{z}_{0,sp}^{2}$ for PoI regions and reuse tokens $\hat{\tilde{z}}_0^{1}$ from \(\mathcal{M}_a\) for non-PoI regions to construct tokens $\tilde{z}_0^{2}$.
%
For feature maps $f_3$ and $f_4$, we reuse tokens from memory token pool $\mathcal{M}_b$ as the remaining positions of the current frame.
%
These memory tokens are inserted into $z_L^{t}$, forming a complete token sequence $z_F^{t}$ of length $N$.
%
The MFR is responsible for constructing a complete feature map $f_3$ from $z_F^{t}$. 
%
The remaining computations are similar to Eq.~\ref{deconv} and Eq.~\ref{conv}
\end{comment}

In the end of inference phase, memory token pools $\mathcal{M}_a$ and $\mathcal{M}_b$ are updated by
\begin{equation}
\mathcal{M}_a \leftarrow \tilde{z}_{0}^{k}, \mathcal{M}_b\leftarrow z_L^{k}.
\end{equation}
%This approach always keeps the number of tokens in the two memory token pools as \(N\).

\subsection{Probability-based Patch Sampling}\label{subsec_III}

\begin{figure}[!t]
\begin{center}
\includegraphics[width=1\linewidth]{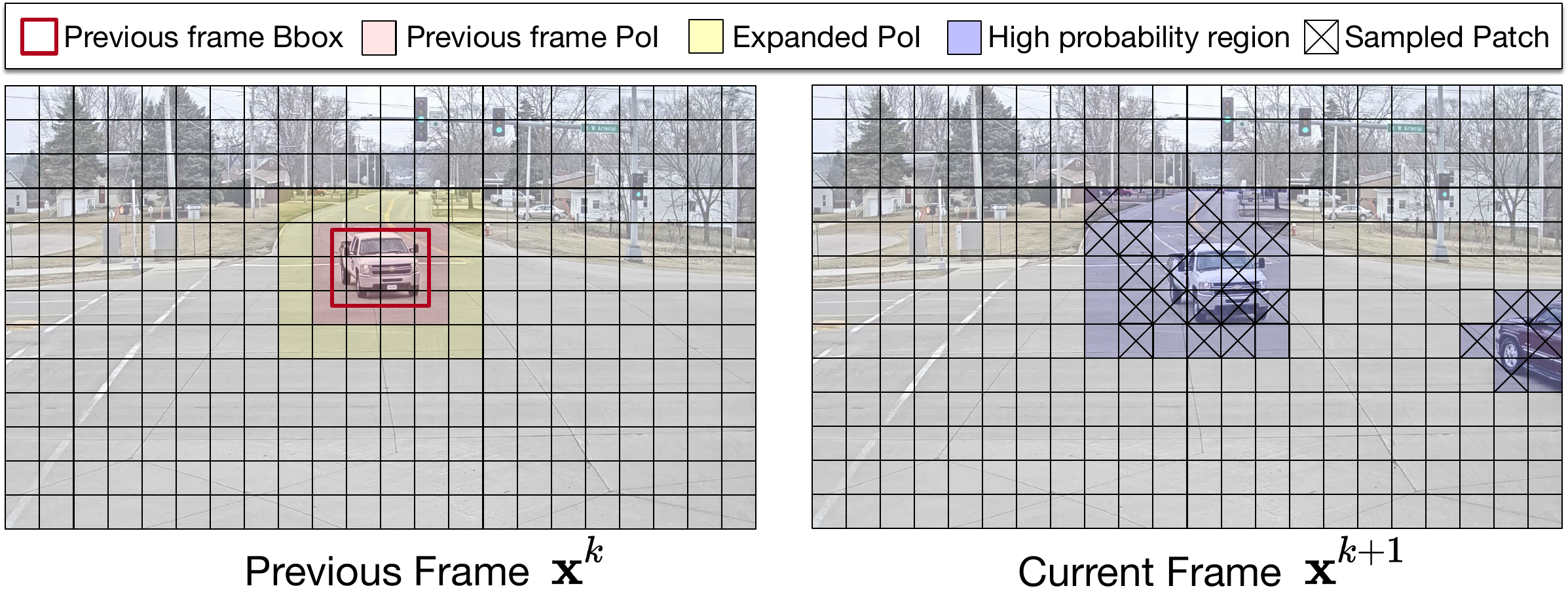}
\caption{Our probability-based patch sampling mechanism. The blue patch indicates that it is highly likely to be determined as an PoI. Zoom in for the best view.}
\label{fig:pps}
\end{center}
\end{figure}

\begin{algorithm}[!t]
% \SetAlgoLined
% \SetKwFunction{Bbox2PoI}{Bbox2PoI}
\KwIn{Previous frame $\mathbf{x}^k$, current frame $\mathbf{x}^{k+1}$, pixel diffrences thereshold $F$, sampling rate $p$, expanded bbox margin $m$, bounding boxes $B^k=\{[\mu_{1,1}^k,\nu_{1,1}^k,\mu_{2,1}^k,\nu_{2,1}^k],...\}$ in previous frame.}
\KwOut{Sampled patches ${R}^k_s$}
$R^{k}_{s} \leftarrow[\emptyset]$	\# initialize the sampled patch list\;
\For{$b \in B^k$}{
$P^k\leftarrow \textit{bbox\_to\_PoI}(b)$\;
$P^k_{s}\leftarrow \textit{expand\_PoI}(P^k,m)$\;
$R^k_s.\textit{append}(P^k_{s})$\;
}
$\mathbf{f}^k\leftarrow \textit{calculate\_pixel\_diff}(\mathbf{x}^k,\mathbf{x}^{k+1})$\;
\For{$f^k_{i} \in \textbf{f}^k$}{
\If{$\textit{sum}(f^k_{i})>F \land f^k_{i}\notin R^k_s$}{
$R^k_s.\textit{append}(f^k_{i})$\;
}
}
${R}^k_s\leftarrow \textit{random\_sample}(R^k_s,p)$\;
% $P^k\leftarrow{Bbox2PoI{(B^k)}}$\;
% $P^k_{s}\leftarrow{Expand\_PoI{(P^k,m)}}$\;
% \textbf{return} $\hat{R}^k_s$\;
\SetKwProg{Fn}{Function}{:}{\KwRet$[\hat{x}_1^k,\hat{y}_1^k,\hat{x}_2^k,\hat{y}_2^k]$}
\Fn{\textit{bbox\_to\_PoI}$(b)$}{
$\mu_1^k\leftarrow{\mu_1^k//16\times{16}}$\;
$\nu_1^k\leftarrow{\nu_1^k//16\times{16}}$\;
$\mu_2^k\leftarrow{(\mu_2^k//16+1)\times{16}}$\;
$\nu_2^k\leftarrow{(\nu_2^k//16+1)\times{16}}$\;    
}
\SetKwProg{Fn}{Function}{:}{\KwRet$[\widetilde{\mu}_1^k,\widetilde{\nu}_1^k,\widetilde{\mu}_2^k,\widetilde{\nu}_2^k]$}
\Fn{\textit{expand\_PoI}$(P^k,m)$}{
$\widetilde{\mu}_1^k\leftarrow{\hat{\mu}_1^k-m\times 16}$\;
$\widetilde{\nu}_1^k\leftarrow{\hat{\nu}_1^k-m\times 16}$\;
$\widetilde{\mu}_2^k\leftarrow{\hat{\mu}_1^k+m\times 16}$\;
$\widetilde{\nu}_2^k\leftarrow{\hat{\nu}_1^k+m\times 16}$\;    
}

\caption{Probability-based Patch Sampling }
\label{alg:PPS}
\end{algorithm}
Determining the PoIs in the next frame is a critical challenge.
Too few patches may cause the model to miss detections, while too many patches could transmit unnecessary background information and add extra computational overhead.
To address this issue, we design a lightweight and effective probability-based patch sampling mechanism that leverages history information to determine the PoIs, shown in Algorithm.~\ref{alg:PPS}.
Our insight is that the region where the object appears in the next frame will be near the bounding box of the previous frame with high probability.

As shown in Figure~\ref{fig:pps}, the two frames are adjacent. 
zIn the first frame $\mathbf{x}^k$, we have obtained the bounding box of the objects (box in red), and all patches occupied by this bounding box (patches in red) are considered as the PoIs of the previous frame.
Next, we expand the PoIs region by appending $m$ (referred to as the expanded bbox margin) rows and columns around the perimeter (Lines 1-6), including the top, bottom, left, and right edges (patches in yellow).
These patches are regarded as the sampling region $R_s^{k}$ for the current frame $\mathbf{x}^{k+1}$ (patches in blue).
We use inter-frame differences to determine the regions that need additional transmission to alleviate the miss of new objects that enter the scene for the first time.
Specifically, we convert $\mathbf{x}^{k}$ and $\mathbf{x}^{k+1}$ to greyscale images and compute the pixel difference $\mathbf{f}^{k}$ between the two greyscale images (Line 7).
Next, we divide the greyscale difference image $\mathbf{f}^{k}$ into non-overlapping $16\times16$ patches ($\mathbf{f}^k = \{{f}^k_{1}, \cdots, f^k_{N}\}$) identifying patches outside the $R_s^{k}$ with pixel differences exceeding the threshold $F$ (Lines 8-12).
The new objects are likely to be found in these patches, which are also marked as sampling regions.
Finally, we randomly sample $p\%$ of patches from the sampling region as PoIs~(Line 13) because our backbone is not sensitive to whether the object area is complete, and the background can be reconstructed using MFR.
%
%For scenes with changing backgrounds, we argue that employing optical flow methods such as flow net and selecting patches with magnitudes above a certain threshold as PoIs is equally effective.

%To explore potential design spaces, we also follow the token pruning strategy~\cite{rao2021dynamicvit} by employing a learnable gate to filter a certain proportion of patches as PoIs on the camera side. 

\subsection{Adaptive Keyframe Interval Switching Algorithms}\label{subsec_IV} 

\begin{algorithm}[t]
% \SetAlgoLined
\KwIn{Optical flow threshold $\beta$, keyframe $\hat{\mathbf{x}}_{r}^{1}$, last frame in the interval $\mathbf{x}_{r}^{K}$, current keyframe interval $K_{r}$, bounding boxes $B_{r}^{1}$ in keyframe.}
\KwOut{Next keyframe interval $K_{r+1}$}
% Initialize the upper bound $K_{upper}$ and lower bound $K_{lower}$ of interval\;
$V_{bbox}\leftarrow 0, A_{bbox}\leftarrow 0$\;
$\mathbf{V}\leftarrow \textit{caculate\_optical\_flow\_diff}(\hat{\mathbf{x}}_{r}^{1},\mathbf{x}_{r}^{K})$\;
$R_b \leftarrow \textit{zeroes}(\mathbf{V}.\textit{row},\mathbf{V}.\textit{column})$ \# initialize a mask matrix of the same shape as $\mathbf{V}$\;
\For{$b \in B_{r}^{1}$}{
$R_b(b)\leftarrow 1$\;
% \For{pixel $\in b$}{
% \If{pixel $\notin R_b$}{
%     $R_b.\textit{append}(\textit{pixel})$
%     }} 
}
$V_{bbox}\leftarrow \textit{sum}(\mathbf{V}(R_b))$ \# sum the magnitudes in the bbox regions\;
$A_{bbox}\leftarrow Area(R_b)$\;
$\overline{V}_{bbox}\leftarrow \frac{V_{bbox}}{A_{bbox}}$\;
\uIf{$\overline{V}_{bbox}<\beta \land K_{r}<K_{upper}$ }{
$K_{r+1}\leftarrow K_{r}+1$
}\ElseIf{$\overline{V}_{bbox}>\beta \land K_{r}>K_{lower}$ }{
$K_{r+1}\leftarrow K_{r}-1$
}
\textbf{return} $K_{r+1}$\;
\caption{Adaptive Keyframe Interval Switching }
\label{alg:akis}
\end{algorithm}

Keyframe interval introduces a trade-off between accuracy and bandwidth, as previously discussed in Sec.~\ref{sec:tradeoff}.
A promising method for adjusting keyframe interval is to assign a smaller value to those ``complex" videos with rapid changes (e.g., dash camera) and a larger value to those ``simple" videos with slower changes (e.g., surveillance camera).
To do this, we propose the AKIS algorithms to control the keyframe interval to achieve optimal accuracy and bandwidth trade-off.
Inspired by the optical flow method~\cite{farneback2003two}, our insight is to set a threshold for optical flow difference values, using it as a bidirectional switch to adjust the keyframe interval.

The AKIS algorithm is designed to determine the length of the r+1-th keyframe interval $K_{r+1}$ based on the optical flow information from the current interval $K_{r}$, shown in Algorithm.~\ref{alg:akis}.
Specifically, we calculate the optical flow map $\mathbf{V}$ between the keyframe $\hat{\mathbf{x}}_r^1$ and the last non-keyframe $\mathbf{x}_r^K$ in the current interval (Line 2).
The magnitude of an optical flow map represents the extent of pixel motion between two frames.
To mitigate the impact of noise (i.e., background pixel, camera shifts) on the optical flow map, we retain only the non-overlapping bounding box regions in keyframes to calculate the magnitude $\mathbf{V}(b)$ (Lines 3-8).
Next, the average optical flow magnitude $\overline{V}_{bbox}$ is calculated as the total magnitude divided by the non-overlapping area of all bounding boxes (Line 9).
Ultimately, a comparison is made between $\overline{V}_{bbox}$ and the threshold $\beta$. 
If $\overline{V}_{bbox}$ is lower than $\beta$, the object changes in the current frame interval are slow, which can be considered a ``simple'' scene.
Consequently, the length of the subsequent interval $K_{r+1}$ is increased by one (Lines 10-11).
On the contrary, if the ``complex'' scene is determined, the length of the next interval $K_{r+1}$ is decreased by one (Lines 12-13).

The algorithm will return a new keyframe interval, thereby enabling the transmission bandwidth to be modified in real time for different scenes.
In scenes where the rate of change for objects or people is minimal, the algorithm can extend the keyframe interval, thereby reducing bandwidth usage between the camera and the edge server while maintaining inference accuracy. 
Conversely, in high-change-rate scenes, the algorithm can continuously decrease the keyframe interval to ensure precision. 
By leveraging the trade-off relationship, this algorithm enables an adaptive video inference system that dynamically balances accuracy and bandwidth.

\subsection{Training Procedure}\label{subsec:train}
We fine-tune the whole detector based on a pre-trained ViT backbone in a video-training manner.
During fine-tuning, for each training batch, we input a pair of adjacent frames \(\{\hat{\mathbf{x}}^k, \mathbf{x}^{k+1}\}\) to the model to simulate our keyframe and non-keyframe inference procedure.
Both frames use identical preprocess pipeline, including data augmentation techniques such as cropping and flipping.
After that, we take the first frame \(\hat{\mathbf{x}}^k\) as the keyframe and feed all patches from  \(\hat{\mathbf{x}}^k\) into the ViT backbone to initialize the memory token pools \(\mathcal{M}_a\) and \(\mathcal{M}_b\). 
Since directly using predicted bounding boxes as results may disturb the proper detection in the next frame during training, we use the ground truth bounding boxes instead of predicted ones from the first frame as the detection results.
Finally, for the second frame \(\mathbf{x}^{k+1}\), we follow Sec.~\ref{subsec_III} to get the PoIs and Sec.~\ref{subsec_II} to conduct a non-keyframe inference.
It's worth noting that we only use the inference results of the second frame for back-propagation and network updates.
In addition, we utilize the pre-trained weights of ViT backbone derived from masked image modeling techniques~\cite{he2022masked}.
%,ren2023tinymim

\section{Implementation}\label{sec_IV}
\begin{figure}[!t]
\begin{center}
\includegraphics[width=1\linewidth]{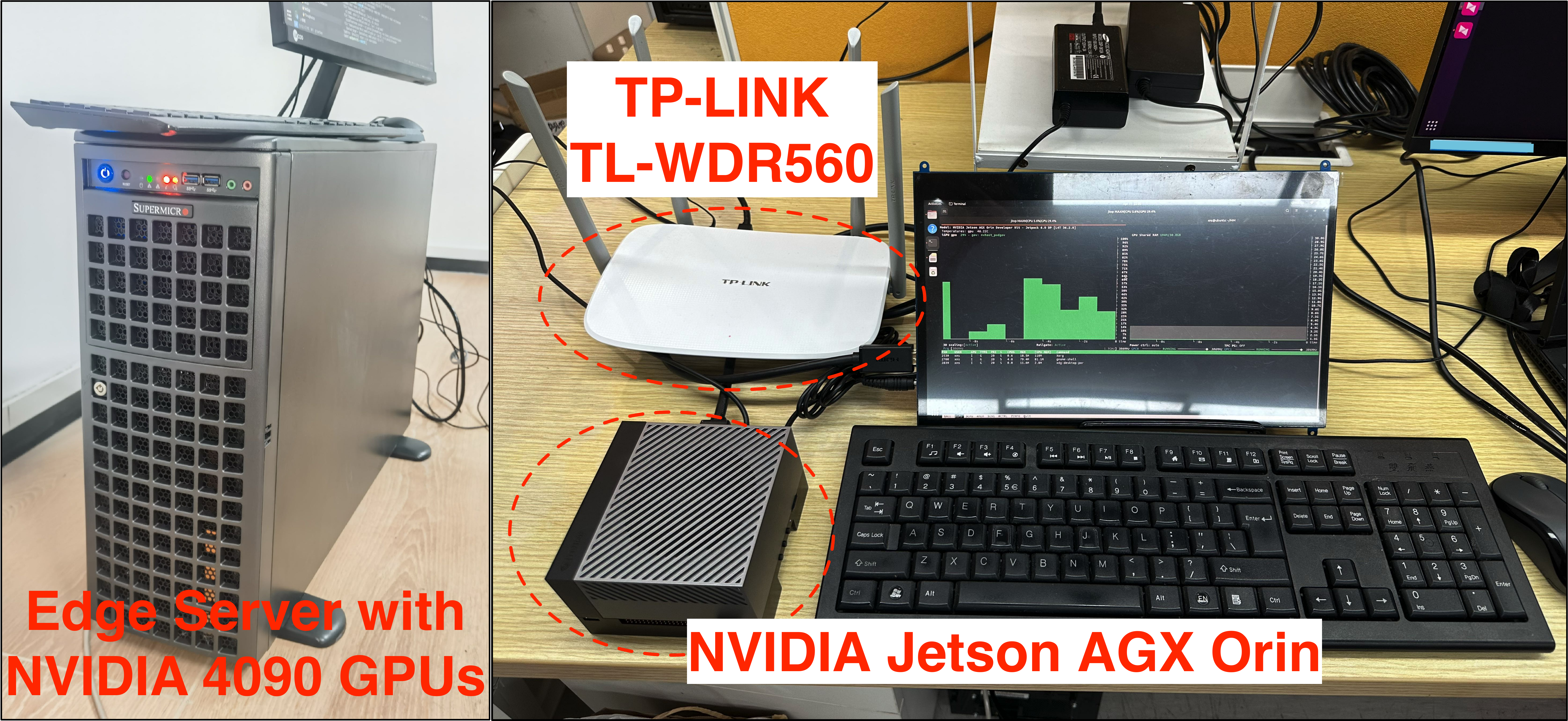}
\caption{Snapshot of our experimental evaluation hardware platform, including an edge server with GPUs and an NVIDIA Jetson AGX as the camera.}
\label{fig:imple}
\end{center}
\end{figure}

We develop a prototype of Arena on an edge server equipped with two Intel(R) Xeon(R) Silver 4310 CPUs, 128 GB of RAM, and two NVIDIA GeForce RTX 4090 GPUs (16384 CUDA Cores), each with 24 GB of VRAM. 
An NVIDIA Jetson AGX Orin running Ubuntu 22.04 LTS serves as the camera.
The edge server and camera are connected with the TP-LINK TL-WDR5620 through a 2.4GHz WiFi network. 
The bandwidth is 93.9 Mbps, measured by iperf.
Arena performs preprocessing on the camera to obtain the PoIs that need to be transmitted.
The camera utilizes the HTTP protocol and creates request sessions to send POST requests to the server.
The server returns the position information for PoIs of the next frame to the camera in JSON format.
The ViT detector is implemented with Python and trained using MMTracking~\cite{mmtrack2020} framework on the edge server. 
%
%All the pre-trained weights are from~\cite{he2022masked,ren2023tinymim}.
%
The PPS is implemented with C++ on the camera. 
Figure~\ref{fig:imple} shows our experimental platform.

\section{Evaluation}\label{sec_V}
In this section, we conduct comprehensive experiments to evaluate the performance of the proposed Arena system. 
We start with the experimental setup and then compare Arena to baseline methods on three metrics.
Next, we give insights into the accuracy and bandwidth trade-off of Arena.
Finally, we report the model analysis.

\subsection{Experimental Setup}
\subsubsection{Performance Metrics} The performance metrics we evaluate include accuracy, bandwidth usage, and end-to-end latency.

$\bullet$ \textbf{Accuracy}: In the object detection task, we evaluate detection accuracy using mAP@0.5, where an object is considered detected if the intersection over union exceeds 0.5. 
This metric is commonly employed in video analytics research.

$\bullet$ \textbf{Bandwidth Usage}: Arena aims to reduce the size of the offloaded data. 
To verify this, we use the data size of the frame sent to the edge server to measure the bandwidth usage. 
Since the bandwidth consumption is different between datasets, we report values after normalization.

$\bullet$ \textbf{End-to-End Latency}: Arena can accelerate the whole video analytics system in multiple aspects, so we compare the average end-to-end latency.
End-to-end latency is defined as the time from the capture of a frame to the completion of inference, including the latency for frame preprocessing on the camera, transmission delay, and inference time.

\subsubsection{Datasets}
We evaluate Arena using the following two datasets. 

$\bullet$ \textbf{MOT17-Det~\cite{milan2016mot16}}: It contains moving or stationary street views with pedestrian bounding box labels from indoor and outdoor settings across 11 distinct regions.
The resolutions of these videos are either $640\times480$ or $1920\times1080$. 
We split the initial 75\% of each video sequence to constitute our training set, with the remaining 25\% serving as the validation set. 

$\bullet$ \textbf{AI City Challenge 2022 (AIC22)~\cite{Naphade22AIC22}}: AIC22 is a real-world multi-camera traffic dataset including 42,683 frames and 88,820 vehicle bounding box labels. 
It contains video footage from a total of 59 surveillance cameras situated at five traffic intersections. 
%
%All the videos have a resolution of 1920×1080. 
%
We select videos from three intersections to constitute our training set while setting the videos from the remaining intersections as the validation set. 

%These two datasets evaluate our system performance across mobile and static camera scenarios.

\subsubsection{Arena's Configurations} 
We employ Faster R-CNN~\cite{ren2015faster} as the two-stage detector framework and ViT-small as the backbone, which features a 12-layer transformer encoder and $D=384$. 
MFR is a one-layer transformer decoder that comprises merely 2.07M (5\% of total) parameters and requires approximately 17.16 (1.9\% of total) GFlops.
%
%The choice of the ViT-small is due to its sufficient compactness to enable real-time detection on standard edge servers.
%
The backbone pre-trained weights are from TinyMiM.
% ~\cite{ren2023tinymim}
%
Unless specifically stated,
% the keyframe interval $K=5$
$p=0.9$, and $F=200$.
For the expanded bbox margin $m$ and 
optical flow threshold $\beta$, we set them to $1$ and $10$ for the MOT17Det dataset, and $3$ and $1.5$ for the AIC22 dataset.

\subsubsection{Baselines}
We compare Arena with the following baselines.

$\bullet$ \textbf{Full-frame Detector (FD)}: 
The camera consistently transmits the full frame to the edge server, which employs a single-frame ViT detector for inference without additional training techniques.

$\bullet$ \textbf{Faster R-CNN Detector~\cite{ren2015faster} (FRD)}: Similar to FD, but the edge server employs a Faster R-CNN detector with ResNet-50.

$\bullet$ \textbf{DDS~\cite{du2020server}}: DDS transmits low-quality videos to the edge server to identify RoIs, which are fed back to the camera. 
The camera subsequently transmits these regions in high quality for second-round inference.
We simulate different qualities by adjusting the resolution.

$\bullet$ \textbf{Reducto~\cite{li2020reducto}}: Reducto filters out frames with minimal impact on the accuracy by comparing low-level video feature differences between two adjacent frames at the camera to reduce bandwidth consumption. 
We employ {\ttfamily Edge} feature and utilize training sets to profile the optimal threshold.

$\bullet$ \textbf{ELF~\cite{zhang2021elf}}: ELF uses attention-based LSTM to predict the bounding box shifts frame by frame. 
It then employs a frame partitioning method to offload the region proposals to multiple edge servers. 
We utilize the official implementation and configure it for three servers with parallel offloading and inference.

\subsection{Overall Performance of Arena}
\begin{figure}[!t]
\begin{center}
\includegraphics[width=1\linewidth]{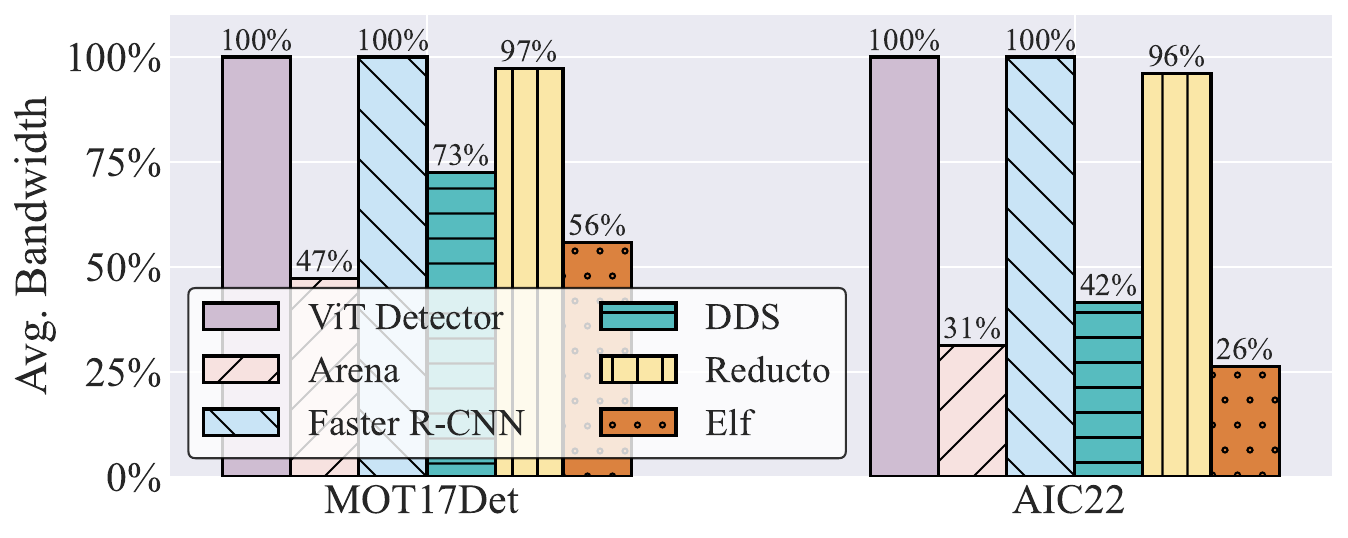}
\caption{The normalized bandwidth usage of different methods on two datasets.}
\label{fig:bandwidth}
\end{center}
\end{figure}

\begin{figure}[!t]
\begin{center}
\includegraphics[width=1\linewidth]{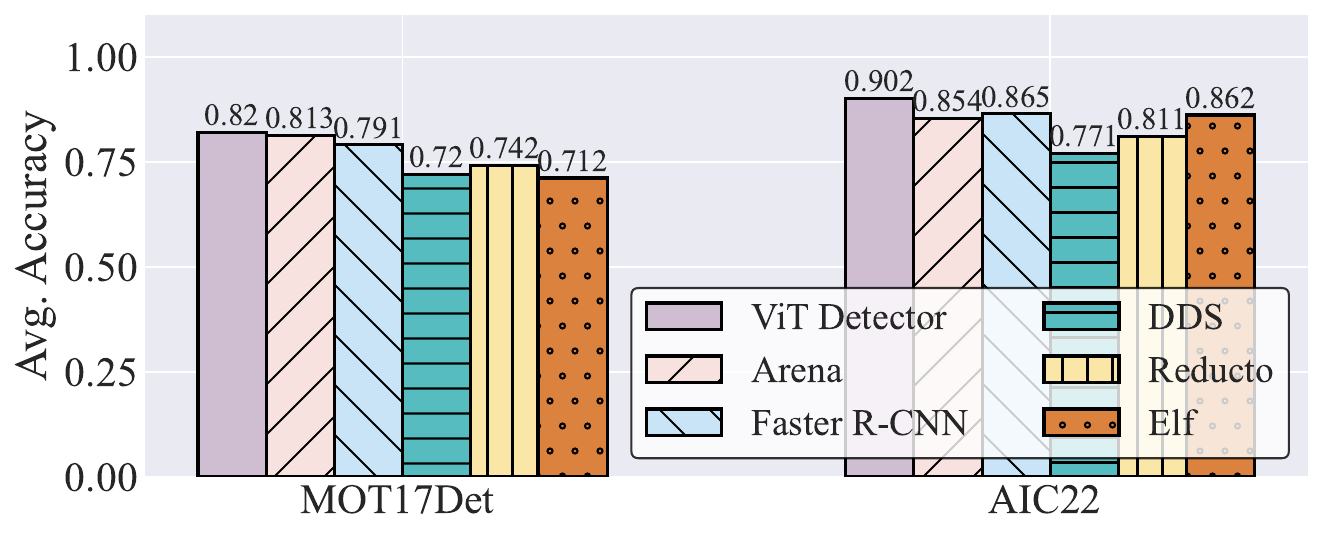}
\caption{The accuracy of different methods on two datasets. Arena can maintain accuracy losses within 1\% and 4\%.}
\label{fig:acc}
\end{center}
\end{figure}

\begin{figure}[!t]
\begin{center}
\includegraphics[width=1\linewidth]{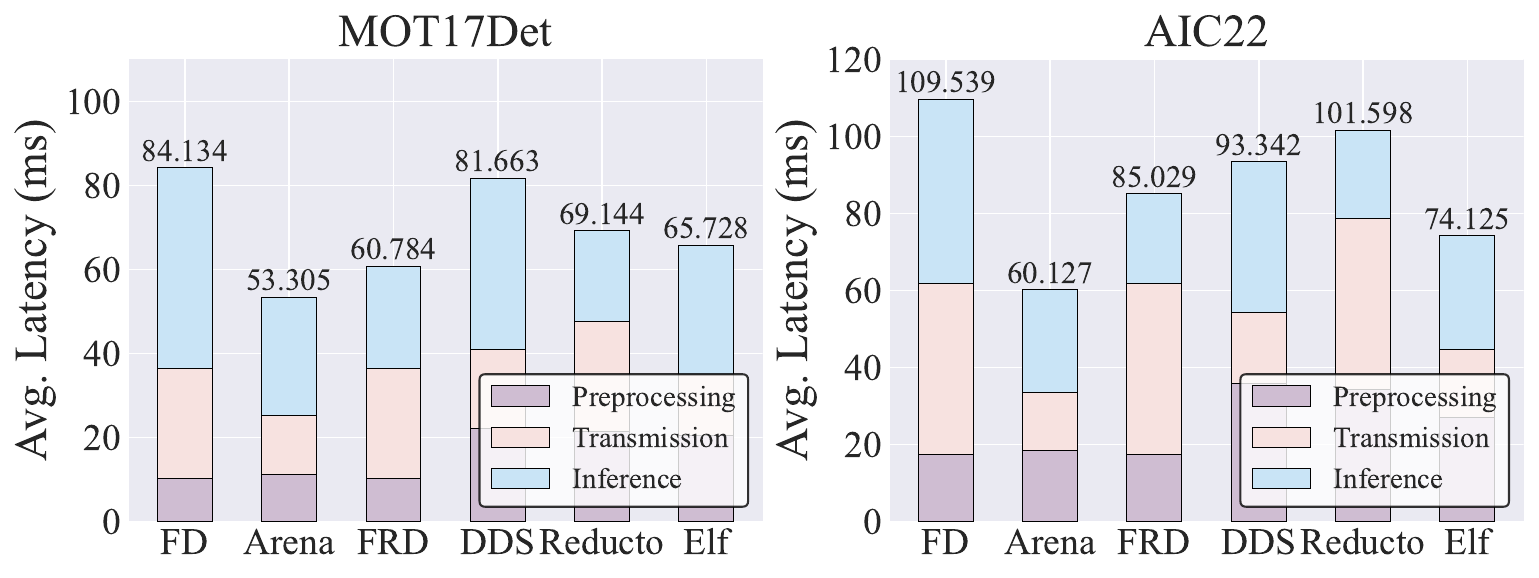}
\caption{The average end-to-end latency per frame of different methods on two datasets. End-to-end latency includes a breakdown of preprocessing, transmission, and inference time.}
\label{fig:latency}
\end{center}
\end{figure}

In Figure~\ref{fig:bandwidth} and Figure~\ref{fig:acc}, we report the normalized bandwidth usage and accuracy of different methods on two datasets. 
Figure~\ref{fig:latency} displays the end-to-end latency of each method, including a breakdown of preprocessing, transmission, and inference time. 
Among these baselines, Arena achieves the best trade-off among bandwidth, latency, and accuracy in both datasets. 
FD and FRD always transmit the full frame for inference, representing the upper bound for bandwidth consumption.
They achieve the highest accuracy on their models but also result in the longest end-to-end latency.
DDS achieves high bandwidth efficiency by transmitting only a few regions in high quality but inevitably incurs additional delay due to two rounds of inference. 
Reducto struggles to adapt to object detection tasks, as removing too many frames significantly reduces its accuracy. 
ELF maintains low bandwidth consumption and end-to-end latency, but this is contingent upon the premise of parallel offloading and inference by three servers. 
In contrast, Arena substantially saves bandwidth usage by pruning patches in non-keyframes, thus reducing both transmission delay and the model inference time.
Meanwhile, Arena utilizes historical frame information to reconstruct features, minimizing the impact on accuracy. 
And Arena integrates AKIS into the end-to-end inference in parallel, which does not affect the end-to-end inference latency.
Specifically, Arena achieves a maximum acceleration of end-to-end latency by $1.58\times$ and $1.82\times$ in two datasets, using only 47\% and 31\% of the bandwidth, while maintaining accuracy losses within 1\% and 5\% as compared to FD, respectively.

\begin{figure*}[!t]
\begin{center}
\includegraphics[width=1\linewidth]{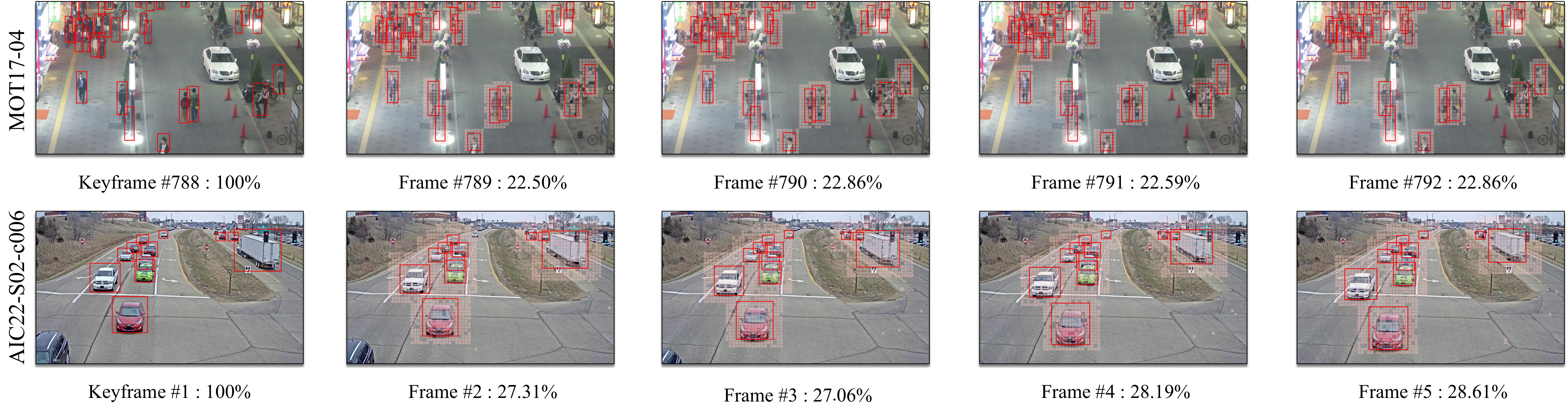}
\caption{Visualization of Arena on two videos. In these two scenes, with a frame interval of 5, $m$ is set to 1 and 3 for MOT17 and AIC22, respectively, $p=0.9$, and $F=200$. Only the red patches in non-keyframe are used for transmission and inference.}
\label{fig:example}
\end{center}
\vspace{-3mm}
\end{figure*}

\begin{figure}[!t]
    \begin{subfigure}{0.23\textwidth}
        % \centering
        \includegraphics[width=.95\linewidth]{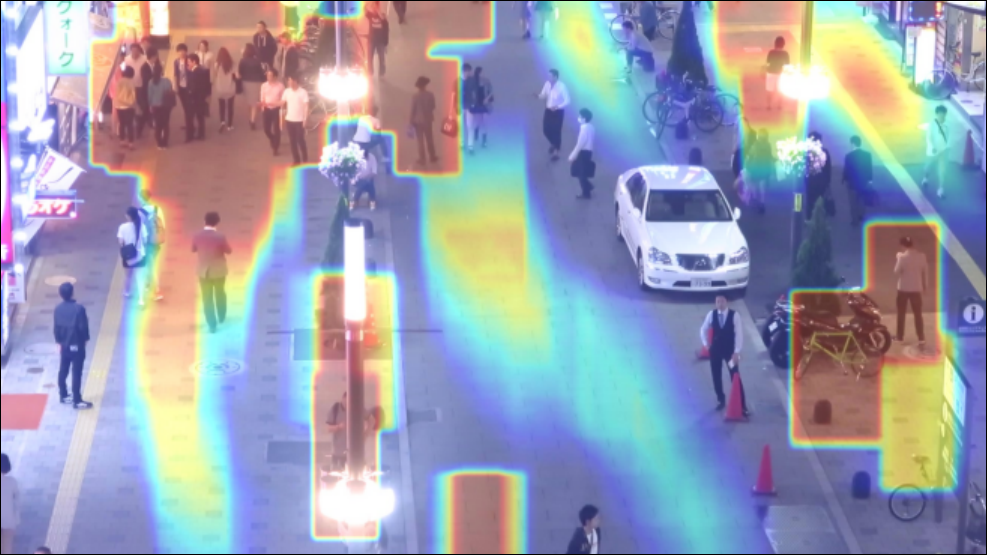}
        \caption{MOT17-04}
        \label{fig1.5:sub1}
    \end{subfigure}
      \begin{subfigure}{0.23\textwidth}
        % \centering
        \includegraphics[width=0.95\linewidth]{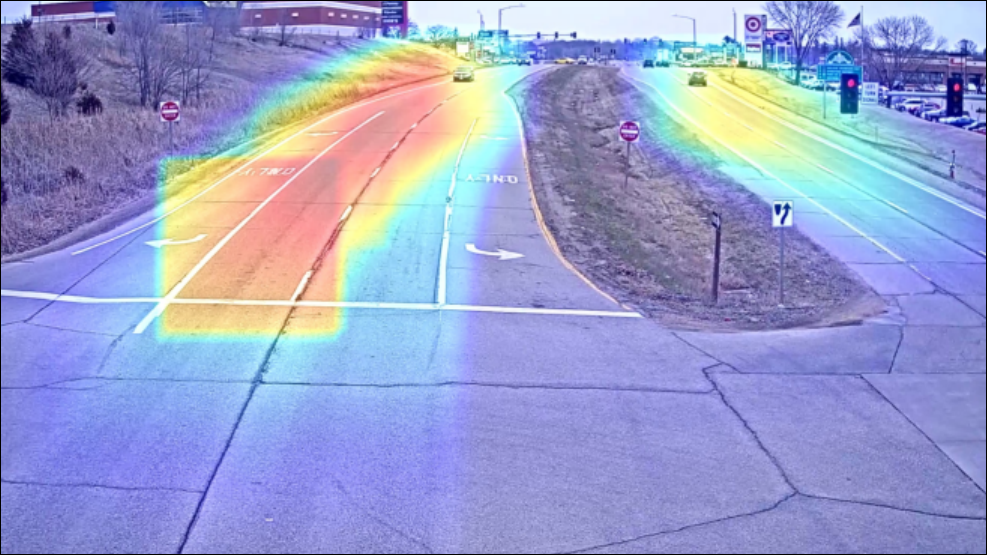}
        \caption{AIC22-S02-c006}
        \label{fig1.5:sub2}
    \end{subfigure}
    \caption{Heatmaps of patches identified as PoIs, where darker areas indicate a higher frequency of offloading to the edge server.}
    \label{fig:heatmap}
\end{figure}

To further illustrate the mechanism of Arena, Figure~\ref{fig:example} presents a visualization of two scenes. 
The first column shows the keyframes, while the remaining four depict non-keyframes. 
In non-keyframes, Arena only samples an average of 22.70\% and 27.79\% of the total patches, yet still achieves satisfactory detection results.
More visualization can be found in the supplementary materials.
Additionally, Figure~\ref{fig:heatmap} displays the heatmaps of the PoIs extracted in these two video sequences, where darker areas indicate a higher frequency of offloading the patch to the edge server.

\begin{figure}[!t]
\begin{center}
\includegraphics[width=1\linewidth]{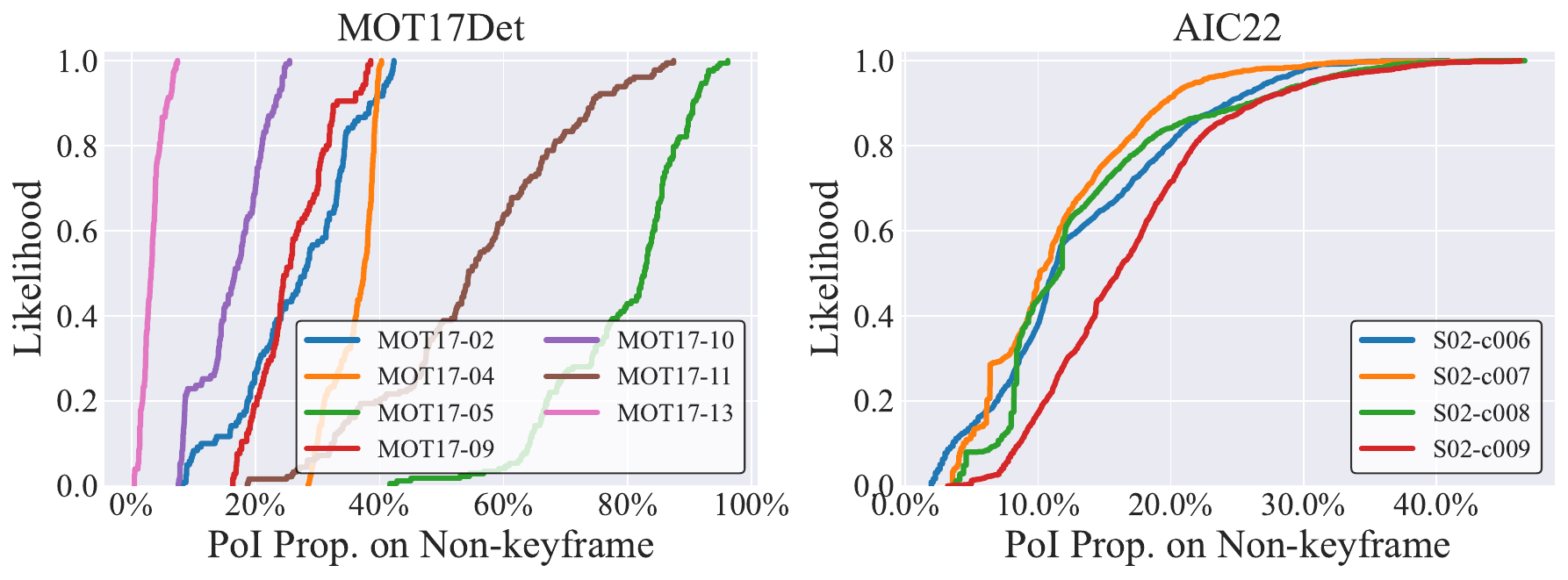}
\caption{The cumulative distribution function (CDF) of PoI proportion of all non-keyframes.}
\label{fig:cdf}
\end{center}
\end{figure}

Moreover, Figure~\ref{fig:cdf} presents the cumulative distribution function (CDF) charts of the proportion of PoI extraction in non-keyframes across various scenes. 
Taking the AIC22 dataset as an example, PoIs are extracted in a more granular and sparse manner, with about 80\% of non-keyframes requiring the transmission of less than 25\% area of the frame.
This is consistent with our findings in the motivation (Sec.~\ref{sec:redundancy}).
The results above reveal that Arena can effectively exploit the extensive redundancy present in the video to achieve end-to-end acceleration.

% \subsection{The performance of AKIS on Arena
% %Accuracy-Bandwidth Usage Trade-off
% }

\begin{figure}[!t]
\begin{center}
\includegraphics[width=1\linewidth]{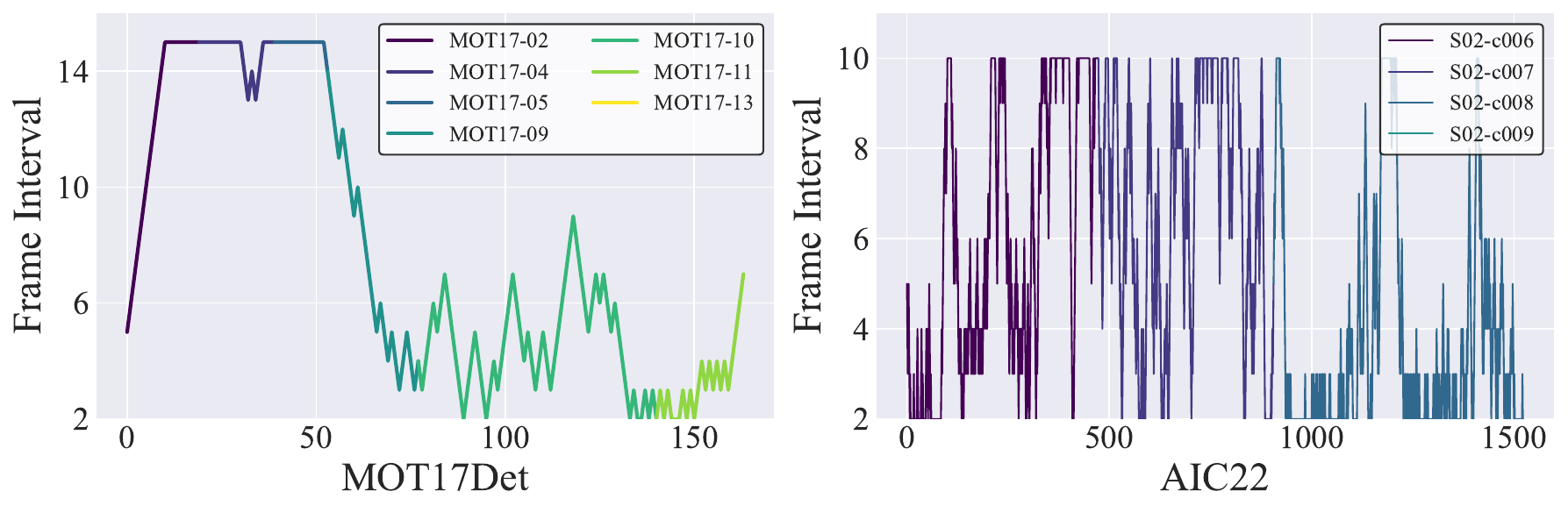}
\caption{Variation of AKIS-based keyframe interval in different scenes in the two datasets.}
\label{fig:AKIS_result}
\end{center}
\vspace{-3mm}
\end{figure}

In Figure~\ref{fig:AKIS_result}, we show the effect of AKIS algorithm on keyframe interval in a variety of scenes, as implemented in Arena.
In the MOT17Det dataset, the upper limit of the keyframe interval is set to 15, and it is evident that AKIS is capable of adapting the keyframe interval in a manner that is conducive to inference in a range of scenes.
For example in certain scenes exhibiting gradual alterations or a limited number of objects—such as ``MOT17-02'' or ``MOT17-04'', AKIS will increase the inference interval in a manner that is adaptive to the circumstances. 
Therefore, AKIS serves to diminish the number of inference operations required for the transmission and inference of complete keyframes, while concurrently ensuring the maintenance of accuracy. 
Furthermore, this strategy enables the reduction of both the inference time and the bandwidth required for the process. 
In more complex scenes, such as ``MOT17-09'' and ``MOT17-11'', AKIS will reduce the keyframe interval in order to ensure accuracy.
A comparable phenomenon can be observed in the AIC22 dataset.
The results demonstrate that the AKIS algorithm, as proposed in Arena, is capable of adaptively modifying the keyframe interval in accordance with scene alterations. 
This enables the attainment of a balance between accuracy and bandwidth in real-time scenes.

\begin{table}[!t]
\caption{The results of whether AKIS is applied to Arena on the two datasets, and the comparison with the fixed keyframe interval.}
\label{tab:AKIS-ablation}
\resizebox{0.48\textwidth}{!}{
\begin{tabular}{@{}cc|cc|cc@{}}
\toprule
\multicolumn{2}{c|}{Datasets} & \multicolumn{2}{c|}{MOT17Det} & \multicolumn{2}{c}{AIC22} \\ \midrule
\multicolumn{1}{c|}{AKIS} & Keyframe Interval & \multicolumn{1}{c|}{mAP@0.5} & BandWidth Usage & \multicolumn{1}{c|}{mAP@0.5} & BandWidth Usage \\ \midrule
\multicolumn{1}{c|}{Yes} & - & \multicolumn{1}{c|}{0.813} & 47\% & \multicolumn{1}{c|}{0.854} & 31\% \\
\multicolumn{1}{c|}{No} & 1 & \multicolumn{1}{c|}{0.82} & 100\% & \multicolumn{1}{c|}{0.902} & 100\% \\
\multicolumn{1}{c|}{No} & 5 & \multicolumn{1}{c|}{0.810} & 53.5\% & \multicolumn{1}{c|}{0.853} & 34\% \\
\multicolumn{1}{c|}{No} & 10 & \multicolumn{1}{c|}{0.804} & 47.3\% & \multicolumn{1}{c|}{0.829} & 25.4\% \\ \bottomrule
\end{tabular}}
\end{table}

\begin{figure}[!t]
\begin{center}
\includegraphics[width=1\linewidth]{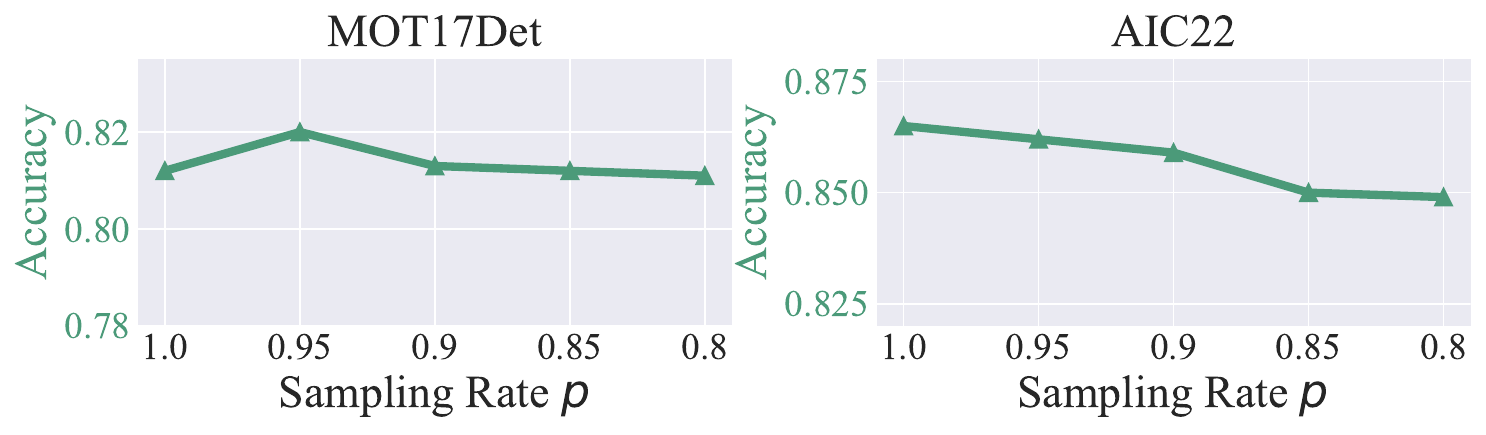}
\caption{The impact of different sampling rates on accuracy.}
\label{fig:p}
\end{center}
\end{figure}

\subsection{Ablation Study}
In order to compose Arena, we propose several optimization adjustments for accuracy and bandwidth,
1) AKIS algorithm, 2) PPS mechanism.
We conduct ablation studies on these modulers to demonstrate their effectiveness for Arena.

\textbf{AKIS algorithm.}
We first compared the results on both datasets with and without the AKIS algorithm implemented in Arena.
In Table~\ref{tab:AKIS-ablation}, compared to the keyframe interval fixed as 5, the AKIS algorithm reduces bandwidth usage dramaticly while maintaining a slight increase in accuracy.
In consequence, it can be demonstrated that the AKIS algorithm can be effectively deployed in Arena, ensuring reconciliation of accuracy and bandwidth consumption while allowing for adaptive adjustment of the keyframe interval based on the complexity of the scenes.

\textbf{PPS mechanism.}
We illustrate the impact of sampling rates within high probability regions on accuracy in Figure~\ref{fig:p}. 
For MOT17Det, accuracy remains relatively stable when the probability $p$ is within the range of 0.8 to 1.0.  
Similarly, for AIC22, a $p$ value exceeding 0.9 consistently keeps high accuracy.
This indicates Arena's strong capability to compensate for the input.
Concurrently, in comparison to the transmission of the entire PoI region ($p=1$), the PPS mechanism is capable of effectively reducing the consumption of transmission bandwidth within the overall video stream, while maintaining accuracy.

\subsection{Model Analysis}
In this subsection, we compare the model complexity of using an ViT detector with other commonly used detectors. 
As shown in Table~\ref{tab:mc}, we benchmark the GMACs and GFlops at a resolution of $800\times1333$ using calflops\footnote{\url{https://github.com/MrYxJ/calculate-flops.pytorch}} and report the frame per second (FPS).
ViT-small is employed as the backbone instead of the ViT-base version in Arena because it can achieve better real-time performance while already maintaining good accuracy.
When processing the videos in the dataset, Arena operates with a computing load below 25$^\triangle$\% or 50$^\triangle$\% versions in the non-keyframe inference phase.
This is because the proportion of PoI in most scenes is below 50\% and even 25\% as shown in Figure~\ref{fig:cdf}.

Finally, we report the model precision in Table~\ref{tab:mp}. 
Though Faster R-CNN offers the fastest inference speed, it requires the transmission of full frames, thus slowing down the end-to-end latency. 
Other models~\cite{zhu2021deformable,zhang2022dino} using the detection transformer~\cite{carion2020end} architecture, despite achieving commendable accuracy, fall short in real-time performance.

\begin{table}[!t]
\caption{Comparison of model complexity with 800 $\times$ 1333 resolution. Here we compare the inference speed (FPS) and complexity (GFlops) between widely used object detectors with different ratios of patch proportion. }
\resizebox{0.48\textwidth}{!}{
\begin{tabular}{@{}c|c|c|c|c|c@{}}
\toprule
\textbf{Model}                                                 & \textbf{Backbone} & \textbf{Params (M)} & \textbf{GMACs} & \textbf{GFlops} & \textbf{FPS} \\ \midrule
ViT Detector (100$^\triangle$\%)                                             & ViT-Small         & 41.453              & 222.955        & 446.333         & 21.85        \\
ViT Detector (75$^\triangle$\%)                                              & ViT-Small         & 41.453              & 200.761        & 401.877         & 26.95        \\
ViT Detector (50$^\triangle$\%)                                              & ViT-Small         & 41.453              & 178.589        & 357.463         & 35.26        \\
ViT Detector (25$^\triangle$\%)                                              & ViT-Small         & 41.453              & 156.395        & 313.007         & 40.82        \\ \midrule
Faster R-CNN~\cite{ren2015faster}        & ResNet50          & 41.123              & 214.648        & 430.087         & 47.47        \\
Deformable-DETR~\cite{zhu2021deformable} & ResNet50          & 39.823              & 203.641        & 408.499         & 2.73         \\
DINO~\cite{zhang2022dino}                & ResNet50          & 47.265              & 286.45         & 574.363         & 2.07         \\ \bottomrule
\end{tabular}}
\begin{tablenotes}
\footnotesize
\item $\triangle$ The proportion of patches used for inference.  %此处加入
%此处加入注释**信息
\end{tablenotes}
\label{tab:mc}
\end{table}

\begin{table}[!t]
\caption{Model Precision}
\resizebox{0.48\textwidth}{!}{
\begin{tabular}{@{}cc|cc|cc@{}}
\toprule
\textbf{} & \textbf{} & \multicolumn{2}{c|}{\textbf{MOT17}} & \multicolumn{2}{c}{\textbf{AIC22}} \\ \midrule
\multicolumn{1}{c|}{\textbf{Model}} & \textbf{Backbone} & \multicolumn{1}{c|}{\textbf{mAP@0.5}} & \textbf{Recall} & \multicolumn{1}{c|}{\textbf{mAP@0.5}} & \textbf{Recall} \\ \midrule
\multicolumn{1}{c|}{ViT Detector} & ViT-Small & \multicolumn{1}{c|}{0.820} & 0.582 & \multicolumn{1}{c|}{0.902} &  0.651\\
\multicolumn{1}{c|}{Faster R-CNN~\cite{ren2015faster}} & ResNet50 & \multicolumn{1}{c|}{0.791} & 0.587 & \multicolumn{1}{c|}{0.865} &  0.620\\
\multicolumn{1}{c|}{Deformable-DETR~\cite{zhu2021deformable}} & ResNet50 & \multicolumn{1}{c|}{0.805} & 0.630 & \multicolumn{1}{c|}{0.880} &  0.648\\
\multicolumn{1}{c|}{DINO~\cite{zhang2022dino}} & ResNet50 & \multicolumn{1}{c|}{0.819} & 0.632 & \multicolumn{1}{c|}{0.876} & 0.656\\ \bottomrule
\end{tabular}}
\label{tab:mp}
\vspace{-3mm}
\end{table}

\section{Related Work}\label{sec_VI}
In this section, we briefly review the latest work in video analytics and also discuss efforts related to accelerating inference with Vision Transformers.

\subsection{Real-Time Video Analytics Systems}
Real-time video analytics is a computationally intensive task. 
Substantial effort~\cite{du2020server,zeng2020distream,lv2023feedback,liu2022adamask,yang2022edgeduet,zhang2021elf,tchaye2022smartfilter,liu2019edge,yang2023javp,yi2020eagleeye} focuses on enhancing the real-time performance of systems in various ways.
%
%Computational offloading is a common approach. 
%
For example, to reduce the transmission data, server-driven methods~\cite{du2020server,zhang2022batch} allow cameras to send low-quality videos to the edge server. 
The edge server then identifies RoIs and provides the camera with their position feedback. 
In the second transmission, only these RoIs, encoded in high quality, are sent to the server to be processed.
Another line of work~\cite{liu2019edge} adopts a lightweight approach that directly extracts RoIs at the cameras and transmits only these regions to the edge server for inference.
Lv et al.~\cite{lv2023feedback} propose a feedback mechanism for detecting RoIs, which involves extracting RoIs from a frame and stitching them together onto a single image for inference. This approach effectively reduces both bandwidth consumption and latency.
%
% Wang et al.~\cite{wang2022edge} design an adaptive offloading algorithm that offloads targets that cameras cannot accurately identify to edge servers for inference using larger models.
%
Liu et al.~\cite{liu2022adamask} use optical flow to identify regions of interest in adjacent video frames and mask the remaining areas to reduce bandwidth consumption.
Additionally, some cloud-edge collaborative systems~\cite{cao2023edge,wang2022vabus} enhance system efficiency by offloading entire or partial computational tasks to the cloud.
Yang et al.~\cite{yang2023javp} develop a strategy that selects whether to offload computing tasks to the Edge or the Cloud based on the complexity of the video content.
Peng et al.~\cite{peng2024tangram} partition object areas from high-resolution videos into patches and offload them to a serverless platform for batch processing.

However, our work is the first to fully leverage characteristics of vision foundation models and achieve acceleration both on computing and transmission in the video analytics system.

\subsection{Efficient Methods for Vision Transformer}

Utilizing the MSA mechanism in ViT leads to computational complexity that typically scales quadratically with the number of tokens.
Consequently, recent efforts have focused on reducing the number of tokens to improve the efficiency of ViT.
These methods primarily fall into two categories: token pruning~\cite{rao2021dynamicvit,liang2022not,long2023beyond,wei2023joint} and token merging~\cite{marin2023token,zeng2022not}.

%Pruning-based methods fall into two primary categories: dynamic pruning and static pruning.
%
Token pruning adjusts the amount of tokens based on specific methodologies. 
Just like what has been done to CNN-type architecture pruning in the past few years, token pruning is applied to accelerate ViT.
Rao et al.~\cite{rao2021dynamicvit} introduces a trainable prediction module (i.e., MLP) that leverages both local and global token information to create a binary mask, which determines the token count for the next layer.
% On the other hand, static pruning involves pruning tokens based on pre-defined pruning rates.
In order to retain some of the information present in the pruned tokens, which may contain contextual data regarding factors such as location and background, several methods~\cite{liang2022not,long2023beyond,wei2023joint} suggest combining the pruned tokens into a single token.
Liang et al.~\cite{liang2022not} categorize tokens into attentive tokens and inattentive tokens based on attention mechanism scores. 
Inattentive tokens are fused together and jointly inputted to the next layer network along with attentive tokens.
Long et al.~\cite{long2023beyond} simultaneously consider both token diversity and importance by performing mering on inattentive and attentive tokens, thereby enhancing accuracy.
Wei et al.~\cite{wei2023joint} merge the pruned tokens with the most similar attentive tokens to retain their own information.
Most merging-based methods rely heavily on clustering algorithms, such as K-Means~\cite{marin2023token}, K-Medoids, and density-peak clustering with K-Nearest neighbors to cluster and merge tokens.
However, these methods present significant challenges in terms of convergence.
Bolya et al.~\cite{bolya2022token} propose a bipartite soft matching algorithm to cluster tokens and subsequently merge similar tokens, thus making it easier to converge.

These ViT acceleration techniques by token reduction are not suited for Arena. 
First, they cannot reconstruct complete feature maps, which are necessary for dense prediction tasks. 
Second, they require the entire frame as input. 
Our proposed Arena, thinking out of the box, effectively adapts efficient ViT for video analytics with its advantages.

\section{Conclusion}\label{sec_VII}
In this paper, we find visual foundation models like ViT also have a dedicated acceleration mechanism for video analytics.
Then, we design Arena, a novel edge-assisted, real-time inference acceleration system tailored for ViT-based models. 
Arena periodically operates in keyframe inference and non-keyframe inference.
The PPS algorithm innovatively guides the filtering of redundant information in video frames during keyframe inference.
MFR is effectively used to reconstruct unstructured feature maps in non-keyframes.
The AKIS algorithm switches inference strategies based on the content of the video.
This system aims to accelerate video analytics by reducing the number of patches offloaded from cameras to the server and alleviating the computation overhead of the model while still maintaining high accuracy. 
We deploy Arena on the prototype that uses an NVIDIA Jetson as a camera device and an Ubuntu desktop as an edge server running real video analytics workloads.
Our results show that Arena can accelerate the inference up to $1.58\times$ and $1.82\times$ on average by only using 47\% and 31\% bandwidth while maintaining a high inference accuracy. 
In the future, we will extend Arena to a broader range of downstream tasks.
% loss within 4\%.

% \begin{acks}
% \end{acks}


\bibliographystyle{IEEEtran}
\bibliography{sample-base}

\end{document}